\documentclass[pre,aps,pacs,twocolumn,eqsecnum]{revtex4}
\usepackage{epsfig}
\usepackage{bm}
\usepackage{color}
\newcommand{\B}[1]{{\bm{#1}}}
\newcommand{\C}[1]{{\mathcal{#1}}}
\usepackage[latin1]{inputenc}
\newcommand{\beq}{\begin{equation}}
\newcommand{\eeq}{\end{equation}}
\newcommand{\bea}{\begin{eqnarray}}
\newcommand{\eea}{\end{eqnarray}}

\begin{document}
\title{The Stability of an Expanding Circular Cavity and the Failure of Amorphous Solids}
\author{Eran Bouchbinder$^{1,2}$, Ting-Shek Lo$^{1,3}$, Itamar Procaccia$^1$ and Elad Shtilerman$^1$}
\affiliation{$^1$Dept. of Chemical Physics, Weizmann Institute of Science, Rehovot 76100, Israel,\\
$^2$Racah Institute of Physics, Hebrew University of Jerusalem, Jerusalem 91904, Israel,\\
$^3$Dept. of Physics, The Chinese University of Hong Kong, Shatin, Hong Kong}
\begin{abstract}
Recently, the existence and properties of unbounded cavity modes,
resulting in extensive plastic deformation failure of
two-dimensional sheets of amorphous media, were discussed in the
context of the athermal Shear-Transformation-Zones (STZ) theory.
These modes pertain to perfect circular symmetry of the cavity and
the stress conditions. In this paper we study the shape stability of the
expanding circular cavity against perturbations, in both the
unbounded and the bounded growth regimes (for the latter the
unperturbed theory predicts no catastrophic failure). Since the
unperturbed reference state is time dependent, the linear stability
theory cannot be cast into standard time-independent eigenvalue
analysis. The main results of our study are: (i) sufficiently small perturbations
are stable, (ii) larger perturbations within the formal linear
decomposition may lead to an instability; this dependence on the
magnitude of the perturbations in the linear analysis is a result of
the non-stationarity of the growth, (iii) the stability of the
circular cavity is particularly sensitive to perturbations in the
effective disorder temperature; in this context we highlight the
role of the rate sensitivity of the limiting value of this
effective temperature. Finally we point to the consequences of the
form of the stress-dependence of the rate of STZ transitions. The
present analysis indicates the importance of nonlinear effects that
were not taken into account yet. Furthermore, the analysis
suggests that details of the constitutive relations appearing in the
theory can be constrained by the modes of macroscopic
failure in these amorphous systems.
\end{abstract}
\pacs{PACS number(s):}
\maketitle

\section{Introduction}

Some of the theoretically most fascinating aspects of crack
propagation in amorphous materials are the
instabilities that are observed in well controlled laboratory
experiments \cite{99FM}. Besides some exceptions, (see for example
\cite{93YS, 95ABP, 03BHP} and also \cite{07LBDF, 07BP}), it would be
fair to say that the observed instabilities are still poorly
understood. It is the opinion of the present authors that the reason
for the relative lack of understanding is that the theory of crack
propagation did not treat cracks as moving free boundaries whose
instabilities stem from the dynamics of the free boundary itself.
Instead, ``crack tip dynamics" were replaced by energy balance within the theory of Linear
Elastic Fracture Mechanics \cite{Freund}, together with an ad-hoc ``law'' of
one nature or another as to where a crack is supposed to move.

In principle this undesirable state of affairs can be greatly
improved within the Shear-Transformation-Zones (STZ) theory of
amorphous materials \cite{79Arg, 79AK,98FL, 07BLanP}. This theory treats developing cracks or growing cavities as free boundaries of a material in which both elasticity
and plasticity are taken into account, preserving all the symmetries
and conservation laws that promise a possibly correct theory of
amorphous materials driven out of mechanical equilibrium. This theory in its various
appearances was compared to a number of experiments and simulations (see below),
with a growing confidence that although not final, STZ theory is
developing in the right direction. Indeed, the application of a
highly simplified version of STZ theory to crack propagation
resulted in physically interesting predictions, explaining how
plasticity can intervene in blunting a crack tip and resulting in
velocity selection \cite{06BPP}. The application of the full
fledged theory of STZ to crack propagation is still daunting
(although not impossible) due to the tensorial nature of the theory
and the need to deal with an extremely stiff set of partial
differential equations with a wide range of time-scales and
length-scales involved. For that reason it seemed advantageous to
apply the full theory to a situation in which the symmetries reduce
the problem to an effectively scalar theory; this is the problem of
a circular cavity developing under circular symmetric stress
boundary conditions \cite{07BLLP, 07BLP}. While this problem does not reach the extreme
conditions of stress concentration that characterizes a running
slender crack, it still raises many physical issues that appear also
in cracks, in particular the give-and-take between elasticity and
plasticity, the way stresses are transmitted to moving boundaries
(in apparent excess of the material yield stress) and most
importantly for this paper, the possible existence of dynamical
instabilities of the moving free boundary. This last issue might
also be connected to the difference between ductile
and  brittle behaviors. In the former, a growing cavity is
likely to remain rather smooth, whereas in the latter, one may expect an
instability resulting in the growth of ``fingers'', possibly ending
up being cracks. It is one of the challenges of the present paper to
examine whether the theory may predict a transition, as a function
of material parameters or a constitutive relation, between these two
types of behavior.

Note that we have chosen to study the problem in a purely 2-dimensional geometry; recently
quasi 2-dimensional systems exhibited interesting failure dynamics in laboratory experiments, where
the 3'rd dimension appears irrelevant for the observed phenomena \cite{07LBDF,04SVC}. Our motivation here is however theoretical, to reduce the unnecessary analytic and numerical complications to a minimum and to gain insight as to the main physical effects under the assumption that the thin
third dimension in real systems does not induce a catastrophic change in behavior. When this assumption fails, as it does in some examples c.f. \cite{99FM}, the analysis must be extended
to include the third dimension. This is beyond the scope of this paper.

In Sec. \ref{EBC} we present the equations that describe the problem at hand and specify their  boundary conditions. Particular attention is paid to distinguish between the general Eulerian formulation which is model-independent (Subsec. \ref{general}) and the constitutive relations involving plasticity where the STZ model is explained (Subsec. \ref{STZ}). This section finishes with the presentation of the unperturbed problem, preparing the stage for the linear stability analysis which is discussed in Sec.
\ref{LSA}. In this section we present a general analysis where inertia and elastic compressibility effects are taken into account. In Appendix \ref{QS} we complement the analysis by considering the ``quasistatic" (when the velocity of the boundary is sufficiently small) and incompressible case (when the bulk modulus is sufficiently large) and show that both formulations agree with one another in the relevant range.
The results of the stability analysis are described in detail in Sec. \ref{results} and a few concluding remarks are offered in Sec. \ref{discussion}.

\section{Equations and Boundary Conditions}
\label{EBC}

\subsection{General formulation}
\label{general}

We start by writing down the full set of equations for a general
two-dimensional elasto-viscoplastic material. A basic assertion of this theory is that
plastic strain tensors in such materials are not state variables since their values depend on the entire history of deformation. Thus, one begins by introducing the total rate of deformation tensor
\begin{equation}
\B D^{\rm tot} \equiv \frac{1}{2}\Big[\B \nabla \B v + \left(\B \nabla \B v\right)^T\Big]  \ , \label{D_tot}
\end{equation}
where $\B v(\B r, t)$ is the material velocity at the location $\B
r$ at time $t$ and $T$ denotes here the transpose of a tensor. This type of Eulerian formulation has the enormous
advantage that it disposes of any reference state, allowing free
discussion of small or large deformations. As is required in an Eulerian frame we
employ the full material time derivative for a tensor $\B T$,
\begin{equation}
\frac{{\cal D} \B T}{{\cal D} t} = \frac{\partial \B T}{\partial t}
+ \B v\cdot \B\nabla \B T +\B T \cdot \B \omega - \B \omega\cdot \B
T \ , \label{material}
\end{equation}
where $ \B\omega$ is the spin tensor
\begin{equation}
\B \omega \equiv   \frac{1}{2}\Big[\B \nabla \B v - \left(\B \nabla \B v\right)^T\Big] \ . \label{omega}
\end{equation}
For a scalar or vector quantity $\B V$ the commutation with the spin
tensor vanishes identically. The Eulerian approach allows a natural
formulation of moving free boundary problems; this will be shown to
lead to a significant advance compared to more conventional
treatments.

The plastic rate of deformation tensor $\B D^{pl}$ is introduced by assuming that the total
rate of deformation tensor $\B D^{\rm tot}$ can be written as a sum of a linear
elastic and plastic contributions
\begin{equation}
\B D^{\rm tot}  = \frac{{\cal D} \B \epsilon^{el}}{{\cal D} t}
+\B D^{pl} \label{el_pl} \ .
\end{equation}
We further assume that $\B D^{pl}$ is a traceless tensor, corresponding to incompressible plasticity. All possible
material compressibility effects in our theory are carried by the elastic component of the deformation.
The components of the linear elastic strain tensor $\B \epsilon^{el}$ are related to
the components of stress tensor, whose general form is
\begin{equation}
\sigma_{ij} = -p\delta_{ij} + s_{ij} \ , \quad p=-\frac{1}{2}
\sigma_{kk} \ , \label{sig}
\end{equation}
according to
\begin{equation}
\epsilon^{el}_{ij} = -\frac{p}{2K}\delta_{ij} + \frac{s_{ij}}{2\mu}
\ , \label{linear}
\end{equation}
where $K$ and $\mu$ are the two dimensional bulk  and shear moduli
respectively. The tensor ${\B s}$ is referred to hereafter as the
``deviatoric stress tensor" and $p$ as the pressure. The equations
of motion for the velocity and density are
\begin{eqnarray}
\label{eqmot1} \rho \frac{{\cal D} \B v }{{\cal D} t} &=& \B \nabla\!\cdot\!\B
\sigma = -\B \nabla p+ \B \nabla\!\cdot\! \B s \ , \\
\quad \frac{{\cal D} \rho}{{\cal D} t}  &=& -\rho \B \nabla\!
\cdot\! \B v \ . \label{eqmot2}
\end{eqnarray}

In order to prepare the general set of equations for the analysis of
a circular cavity we rewrite the equations in polar
coordinates. For that aim we write
\begin{equation}
\label{polarO}
\B \nabla = \B e_r \partial_r  +\frac{\B e_\theta}{r} \partial_\theta ,\quad \B v = v_r \B e_r  + v_\theta \B e_\theta \ ,
\end{equation}
where $\B e_r$ and $\B e_\theta$ are unit vectors in the radial and azimuthal directions respectively. These expressions enable us to represent the divergence operator $\B \nabla \cdot$ in the equations of motion and the covariant derivative $\B v \!\cdot\! \B \nabla$ in the material time derivative of vectors and tensors. Some care should be taken in evaluating these differential operators in polar coordinates since the unit vectors themselves vary under differentiation according to
\begin{equation}
\label{unit_vectors}
\partial_r \B e_r=0,\quad \partial_r\B e_\theta=0,\quad \partial_\theta\B e_r=\B e_\theta,\quad \partial_\theta \B e_\theta=-\B e_r \ .
\end{equation}
We then denote $s_{rr}\equiv -s$, $s_{\theta \theta} \equiv s$,
$s_{r\theta}=s_{\theta r} \equiv \tau$ and using Eqs. (\ref{sig}) we obtain
\begin{eqnarray}
\label{sig_p_s}
\sigma_{rr}  &=& -s -p \ ,\nonumber\\
\sigma_{\theta \theta} &=& s-p \ ,\nonumber\\
\sigma_{r \theta} &=& \sigma_{ \theta r} =\tau \ .
\end{eqnarray}
In this notation the equations of motion
(\ref{eqmot1}) can be rewritten explicitly as
\begin{widetext}
\begin{eqnarray}
\rho \left(\frac{\partial v_r}{\partial t} \!+\! v_r \frac{\partial
v_r}{\partial r}\!+\! \frac{v_\theta}{r} \frac{\partial v_r}{\partial
\theta}-\frac{v_\theta^2}{r}\right)\!&=&\! \frac{1}{r}\frac{\partial \tau}{\partial \theta}
-\frac{1}{r^2} \frac{\partial }{\partial r} \left ( r^2 s \right)\! -\!
\frac{\partial
p}{\partial r} \nonumber  \ , \\
\rho \left(\frac{\partial v_\theta}{\partial t} \!+\! v_r
\frac{\partial v_\theta}{\partial r}\!+\! \frac{v_\theta}{r}
\frac{\partial v_\theta}{\partial \theta} +\frac{v_\theta v_r}{r}\right)\!&=&\!\frac{\partial
\tau}{\partial r}\! +\! \frac{1}{r} \frac{
\partial  s}{\partial \theta}\! -\! \frac{1}{r}
\frac{ \partial  p} {\partial \theta} \!+\!\frac{2 \tau}{r} \ ,  \nonumber\\
\label{EOM}
\end{eqnarray}
\end{widetext}
where $\B \nabla\!\cdot\!\B \sigma$ is calculated explicitly in Appendix \ref{polar}.

Equations (\ref{el_pl}) can be rewritten in components form as
\begin{eqnarray}
\label{eq:DA_ij}
D^{\rm tot}_{ij}&=&
\frac{\partial \epsilon^{el}_{ij}}{\partial t} + \left(\B v \cdot \B \nabla \B \epsilon^{el} \right)_{ij}\\
&+&\epsilon^{el}_{ir}\omega_{rj}+\epsilon^{el}_{i\theta} \omega_{\theta j}-
\omega_{ir}\epsilon^{el}_{rj}-\omega_{i \theta}\epsilon^{el}_{\theta j}+ D^{pl}_{ij}\ .\nonumber
\end{eqnarray}
Here the components of the total rate of deformation tensor are related to the
velocity according to Eqs. (\ref{D_tot}) as
\begin{eqnarray}
D_{rr}^{\rm tot} &\equiv& \frac{\partial v_r}{\partial r},\quad
 D_{\theta\theta}^{\rm tot} \equiv \frac{\partial_\theta v_\theta +
 v_r}{r} \ ,\nonumber\\
D_{r \theta}^{\rm tot} &\equiv& \frac{1}{2} \left[ \partial_r
v_\theta + \frac{\partial_\theta v_r - v_\theta}{r} \right] \ ,
\label{eq:totalrate}
\end{eqnarray}
where the components of the spin tensor $\B \omega$ in Eq. (\ref{omega}) are given by
\begin{eqnarray}
\omega_{rr}&=&\omega_{\theta \theta}=0  \ ,\nonumber\\
\omega_{r \theta}&=& - \omega_{\theta r} = \frac{1}{2} \left[
\frac{\partial_{\theta}v_r -v_\theta}{r} - \partial_r v_\theta
\right] \ .
\end{eqnarray}
The calculation of the tensor $\B v\! \cdot\! \B \nabla \B \epsilon^{el}$ is presented in Appendix \ref{polar}; the linear elastic strain components of Eqs. (\ref{linear}) are given by
\begin{eqnarray}
\epsilon_{rr}^{el} &=& - \frac{p}{2K}  -
\frac{s}{2\mu}\ , \nonumber\\
\epsilon_{\theta \theta}^{el} &=&  - \frac{p}{2K}
+ \frac{s}{2\mu}\ , \nonumber\\
\epsilon_{r \theta}^{el} &=& \epsilon_{\theta r}^{el}=\frac{\tau}{2 \mu} \ .
\label{eq:stress-strain}
\end{eqnarray}

Since most of the materials of interest have a large bulk modulus
$K$, i.e. they are almost incompressible, we assume that the density
is constant in space and time
\begin{equation}
\rho(\B r,t) \simeq \rho \label{density} \ .
\end{equation}
Therefore, Eq. (\ref{eqmot2}) is omitted. Finally, the existence of a free boundary is introduced as the following boundary conditions
\begin{equation}
\sigma_{ij}n_j=0 \ , \label{stressBC}
\end{equation}
where $\B n$ is the unit normal vector at the free boundary.

\subsection{Viscoplastic constitutive equations:\\ The athermal STZ theory}
\label{STZ}

Up to now we have considered mainly symmetries and conservation
laws. A general theoretical framework for the elasto-viscoplastic
deformation dynamics of amorphous solids should be supplemented with
constitutive equations relating the plastic rate of deformation
tensor $\B D^{pl}$ to the stress and possibly to other internal
state fields. We use the constitutive equations of the recently proposed athermal Shear Transformation Zones (STZ) theory \cite{07BLanP}. This theory is based on identifying the
internal state fields that control plastic deformation. The basic
observation is that stressing a disordered solid results in localized reorganizations
of groups of particles. These reorganizations occur
upon surpassing a local shear threshold, and when they involve a finite irreversible shear in
a given direction, we refer to them as an ``STZ transition". Once transformed, due to a local redistribution
of stresses, the same local region resists further deformation in that direction,
but is particulary sensitive to shearing transformation if the local
applied stress reverses its direction. Thus an STZ transition is conceived
as a deformation unit that can undergo configurational
rearrangements in response to driving forces. Furthermore, the
stress redistribution that accompanies an STZ transition can induce the creation and annihilation of other local particle arrangements that  can undergo further localized transitions;  these
arrangements are formed or annihilated at a rate
proportional to the local energy dissipation (recall
that thermal fluctuations are assumed to be absent or negligible). In this sense the interesting localized
events need not depend on ``pre-existing" defects in the material, but can appear and disappear
dynamically in a manner that we describe mathematically next.

This picture is cast into a mathematical form in terms of a scalar field $\Lambda$
that represents the normalized density of regions that can undergo STZ transitions, a tensor $\B m$ that represents the difference between the density of regions that can undergo a transition under
a given stress and the reversed one, and an effective
disorder temperature $\chi$ that characterizes the state of
configurational disorder of the solid \cite{04Lan}. The present state of the theory relates
these internal state fields,
along with the deviatoric stress tensor ${\B s}$, to the
plastic rate of deformation tensor $\B D^{pl}$ according to
\begin{eqnarray}
\label{eq:Dpl}
\tau_0 D^{pl}_{ij} \!=\! \epsilon_0 \Lambda \C C(\bar{s})\left(\frac{s_{ij}}{\bar{s}}-m_{ij}\right),\quad\bar{s} \equiv \sqrt{\frac{s_{ij}s_{ij}}{2}}\ .
\end{eqnarray}
This equation represents the dependence of the plastic rate of deformation on the current
stress $s_{ij}$ and the recent history encoded by the internal state tensorial field $\B m$.
This field acts as a back-stress, effectively reducing the local driving force for STZ transitions, up
to the possible state of jamming when the whole parentheses vanishes. The parentheses
provides information about the orientation of the plastic deformation. The function $ \C C(\bar{s})$
determines the magnitude of the effect, and is re-discussed below. The field $\Lambda$ appears
multiplicatively since the rate of plastic deformation must be proportional to the density of STZ.
The second equation describes the dynamics of the internal back stress field
\begin{eqnarray}
\label{eq:m}&&\tau_0\frac{{\cal D} m_{ij} }{{\cal D} t} =
2\frac{\tau_0 D^{pl}_{ij}}{\epsilon_0 \Lambda
}- \Gamma(s_{ij}, m_{ij})m_{ij}\frac{e^{-1/\chi}}{\Lambda} \ ,\nonumber\\
&&\hbox{with}\quad\Gamma(s_{ij}, m_{ij}) = \frac{\tau_0
s_{ij}D^{pl}_{ij}}{\epsilon_0 \Lambda} \ .\label{Gamma}
\end{eqnarray}
This equation captures the dynamical exchange of stability when the material yields to the applied
stress. The equation has a jammed fixed point when the plastic deformation vanishes, in agreement
with the state of STZ being all in one orientation, without the production of a sufficient number
of new ones in the other orientation. The jammed state is realized when the applied stress
is below the yield stress. When the stress exceeds the threshold value the stable fixed point
of this equation corresponds to a solution with non-vanishing plastic rate of deformation. This
state corresponds to a situation where enough STZ are being created per unit time to allow
a persistent plastic flow. The quantity $\Gamma$ represents the rate of STZ production in response to the flow $\B D^{pl}$.
The next equation, for the density of STZ $\Lambda$, is an elementary fixed point equation reading
\begin{equation}
\label{eq:Lambda} \tau_0 \frac{{\cal D} \Lambda }{{\cal D} t} =
\Gamma(s_{ij}, m_{ij})\left(e^{-1/\chi}- \Lambda\right) \ .
\end{equation}
The unique fixed point of this equation is the equilibrium solution $\Lambda=e^{-1/\chi}$ where
$\chi$ is a normalized temperature-like field which is not necessarily the bath temperature
when the system is out of thermal and/or mechanical equilibrium. The last equation is
for this variable, reading
\begin{eqnarray}
\label{eq:chi} \tau_0 c_0 \frac{{\cal D} \chi }{{\cal D} t}
&=& \epsilon_0 \Lambda \Gamma(s_{ij},
m_{ij})\left[\chi_\infty\left(\tau_0\bar{D}^{pl}\right)-\chi\right],\nonumber\\
\hbox{with}\quad \bar{D}^{pl}&\equiv & \sqrt{\frac{D^{pl}_{ij}D^{pl}_{ij}}{2}} \ .
\end{eqnarray}
This is a heat-like equation for the configurational degrees of freedom; it is discussed in detail below.
Here and elsewhere we assume that quantities of stress dimension
are always normalized by the yield stress $s_y$; this is
justified as the STZ equations exhibit an exchange of dynamic
stability from jamming to flow at $s\!=\!1$, i.e. at a stress that
equals to $s_y$ \cite{07BLanP}. The set of Eqs. (\ref{eq:Dpl})-(\ref{eq:chi}) is a tensorial
generalization of the effectively scalar equations derived in
\cite{07BLanP}; such a generalization can be obtained by following the
procedure described in Ref. \cite{05Pech}. In these equations,
$\tau_0$ is the elementary time scale of plasticity, $\epsilon_0$ is
a dimensionless constant and
$c_0$ is a specific heat in units of $k_B$ per particle.

A weak point of the theory is the lack of a first-principle derivation that determines  the
function ${\cal C}(s)$ in Eq. (\ref{eq:Dpl}), which lumps together
much of the microscopic physics that controls the stress-dependent
rate of STZ transitions. Our theory constrains it to
be a symmetric function of $s$ that vanishes with vanishing
derivatives at $s\!=\!0$, due to the athermal condition that states
that no transitions can occur in a direction opposite to the
direction of $s$ \cite{07BLanP}. This constraint is not sufficient, however, to
determine ${\cal C}(s)$. To appreciate the uncertainties, recall that
STZ transitions are relaxation events, where energy and stress are
expected to re-distribute. Even without external mechanical forcing, aging in
glassy systems involves relaxation events that are poorly
understood \cite{01LN}. The situation is even more uncertain
when we deal with dynamics far from mechanical equilibrium. The best one can do at present is to choose the function ${\C C}(s)$ by examining its influence on the resulting
macroscopic behaviors \cite{07BL}.
Thus in this paper we will examine the sensitivity of the stability
of the expanding cavity to two different choices of ${\cal C}(s)$. At
present we use the one-parameter family of functions, ${\cal
C}(\bar{s})=\C F(\bar{s}; \zeta)$, proposed in \cite{07BLanP}
\begin{equation}
\label{C_s} \C F(\bar{s};\zeta)\equiv
\,\frac{\zeta^{\zeta+1}}{\zeta!}\int_0^{|\bar
s|}(|\bar s|-s_{\alpha})\,s_{\alpha}^{\zeta}\,\exp (-\zeta \,
s_{\alpha})\,d s_{\alpha}\ .
\end{equation}
The integral is over a distribution of transition thresholds whose
width is controlled by a parameter $\zeta$ (and see \cite{07BLanP} for
details). For finite values of $\zeta$ there can be nonzero
sub-yield plastic deformation for $|s|\!<\!1$. This behavior is well
documented in the literature cf. \cite{Lubliner} in the context of experimental stress-strain
relations and plastic deformations. We note that for $s$
very small or very large,
\begin{eqnarray}
\label{limits}
{\cal C}(s) &\sim&  s^{\zeta+2} \quad\hbox{for}\quad  s \to 0^+ \ ,\nonumber\\
{\cal C}(s) &\simeq& s-1 \quad\hbox{for}\quad  s \gg 1
\ .
\end{eqnarray}
In Sec. \ref{rate_function} we propose a different one-parameter family of functions $\C G(\bar{s}; \lambda)$ and study in detail the implications of this different choice on the stability of the expanding cavity.

\begin{figure}
\centering \epsfig{width=.47\textwidth,file=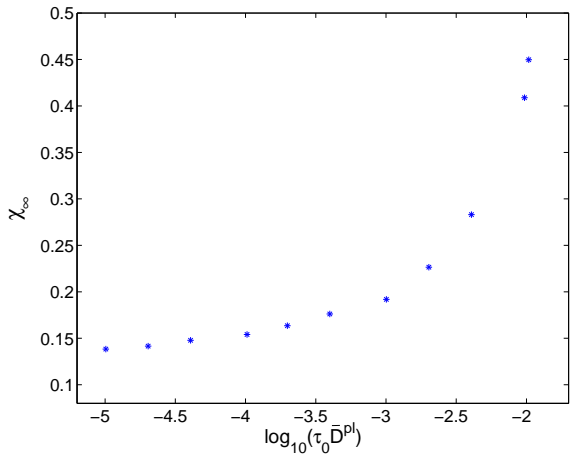}
\caption{(Color online) A typical relation between $\chi_\infty$ and
$\log_{10}\!\!\left(\tau_0\bar{D}^{pl}\right)$ for a temperature
significantly smaller than the glass transition temperature. Data courtesy of T. Haxton and A. Liu. Note that the data were scaled properly.
}\label{HL}
\end{figure}

Eq. (\ref{eq:chi}) deserves special attention. It is a heat-like equation
for the effective disorder temperature $\chi$ with a fixed-point
$\chi_\infty$ which is attained under steady state
deformation. This reflects the observations of Ref. \cite{Ono}, where the effective temperature $\chi$
was shown to attain a unique value in the limit
$t_0\bar{D}^{pl}\!\to\! 0$, where $t_0$ was the particles
vibrational time scale. Indeed, in most applications, realistically {\em
imposed} inverse strain rates are much larger than the elementary time
scale $t_0$, i.e. $t_0\bar{D}^{pl}\!\ll\! 1$. If we identify our $\tau_0$
with the vibrational time scale $t_0$ (see for example \cite{07BLanPb}), we
conclude that
$\chi_\infty$ can be taken as a constant, independent of the plastic
rate of deformation. This assumption was adopted in all
previous versions of STZ theory. Note also that a low plastic rate
of deformation is associated with $s\!\to\! 1^+$, i.e. a deviatoric stress
that approaches the yield stress from above. However, the situation
might be very different in free boundary evolution problems, where
high stresses concentrate near the boundary, reaching levels of a few times
the yield stress. Estimating $\chi$ in the typical range of $0.1-0.15$ \cite{07BLanPb, 07SKLF},
$e^{-1/\chi}$ is in the range $10^{-4}\!-\!10^{-3}$. Therefore, estimating
the other factors in Eq. (\ref{eq:Dpl}), for the high stresses near
the free boundary, in the range $1\!-\!10$, we conclude that
$\tau_0\bar{D}^{pl}$ can reach values in the range
$10^{-4}\!-\!10^{-2}$. Very recent simulations \cite{07HL} demonstrated convincingly that in this range
of normalized plastic rates of deformation, $\chi_\infty$ shows a
considerable dependence on this rate, see Fig. \ref{HL}. Since  $\chi$
affects plastic deformation through an exponential Boltzmann-like
factor, even small changes of $\chi_\infty$ in Eq. (\ref{eq:chi})
can generate significant effects \cite{comment0}. This issue is of particular
importance for the question of stability (and localization) under
study since the strain rate sensitivity of $\chi_\infty$ might
incorporate an instability mechanism; fluctuations in the plastic
rate of deformation, caused for example by fluctuations in $\chi$,
can induce, through $\chi_\infty$, a further localized increase in
plastic deformation and so on. This intuitive idea will be studied
in the analysis to follow.

The set of Eqs. (\ref{eq:Dpl})-(\ref{eq:chi}) \cite{comment} (and slight variants)
was shown to capture viscoelastic behavior in a variety of examples. These include small stress and finite plasticity at intermediate stresses \cite{00FL}, a transition to flow at the
yield stress (as discussed above) \cite{07BLanP}, the deformation dynamics of simulated amorphous silicon \cite{07BLanPb}, the necking
instability \cite{03ELP}, the deformation dynamics near stress
concentrations \cite{07BLLP}, the cavitation instability \cite{07BLP} and
strain localization \cite{07MLC}. In this work we focus on the
implications of these constitutive equations on the stability of
propagating free boundaries in relation to the failure modes of
amorphous solids.
\subsection{The unperturbed problem}
\label{zeroth}

In this subsection we adapt the general theory to the circular
symmetry of the unperturbed expanding cavity problem. We consider an
infinite medium with a circular cavity of radius $R^{(0)}(t)$,
loaded by a radially symmetric stress $\sigma^{\infty}$ at infinity.
The superscript $(0)$ in all the quantities denotes the fact that
they correspond to the perfectly symmetric case that is going to be
perturbed later on. For the perfect circular symmetry the velocity
field $\B v^{(0)}(\B r,\theta)$ is purely radial and independent of the azimuthal
angle $\theta$, i.e.
\begin{equation}
v_r^{(0)}(\B r,t)=v_r^{(0)}(r,t),\quad v_\theta^{(0)}(\B r,t) = 0 \
.
\end{equation}
This symmetry also implies that
\begin{equation}
\tau^{(0)}(\B r,t) = 0, \quad {D^{pl}_{r \theta}}^{(0)}(\B r,
t)=0,\quad m^{(0)}_{r \theta}(\B r,t)=0
\end{equation}
and all the diagonal components are independent of $\theta$.
Eqs. (\ref{el_pl}), after a simple manipulation, can be rewritten as
\begin{eqnarray}
\frac{v_r^{(0)}}{r}+\frac{\partial v_r^{(0)}}{\partial r}\! &=&\!
-\frac{1}{K}\left(\frac{\partial p^{(0)}}{\partial t}+v_r^{(0)}
\frac{\partial p^{(0)}}{\partial r}\right)\ , \label{kinematic1a} \\
\frac{v_r^{(0)}}{r}-\frac{\partial v_r^{(0)}}{\partial r}\!&=&\!
\frac{1}{\mu}\left(\frac{\partial s^{(0)}}{\partial t}+v_r^{(0)}
\frac{\partial s^{(0)}}{\partial r}\right)\!+\!2{D^{pl}}^{(0)} \ .
\nonumber\\ \label{kinematic2a}
\end{eqnarray}
where we have defined
\begin{equation}
{D^{pl}_{\theta \theta}}^{(0)}=-{D^{pl}_{rr}}^{(0)} \equiv
{D^{pl}}^{(0)} \ . \label{D}
\end{equation}
The equations of motion (\ref{EOM}) reduce to
\begin{equation}
\rho \left(\frac{\partial v_r^{(0)}}{\partial t} + v_r^{(0)}
\frac{\partial v_r^{(0)}}{\partial r} \right)= -\frac{1}{r^2}
\frac{\partial }{\partial r} \left (r^2 s^{(0)} \right) -
\frac{\partial p^{(0)}}{\partial r}  \ .\label{EOM0}
\end{equation}
The boundary conditions are given by
\begin{eqnarray}
\sigma^{(0)}_{rr}(R^{(0)},t)=-p^{(0)}(R^{(0)},t)-s^{(0)}(R^{(0)},t)=0 \ ,\nonumber\\
\sigma^{(0)}_{rr}(\infty,t)=-p^{(0)}(\infty,t)-s^{(0)}(\infty,t)=\sigma^{\infty}
\label{BC0}.
\end{eqnarray}
The initial conditions are chosen to agree with the solution of the static linear-elastic problem, i.e.
\begin{eqnarray}
p^{(0)}(r,t=0)&=&-\sigma^{\infty}\ ,  \nonumber\\
s^{(0)}(r,t=0)&=& \sigma^{\infty}\frac{\left(R^{(0)}(t=0)\right)^2}{r^2} \ , \nonumber\\
v_r^{(0)}(r,t=0)&=& 0 \label{initial0} \ .
\end{eqnarray}
This choice reflects the separation of time scales between elastic and plastic responses. This
separation of time scales can be written explicitly in terms of the typical elastic wave speed, the
radius of the cavity and the time scale of plasticity:
\begin{equation}
R^{(0)}(t=0)\sqrt{\frac{\rho}{\mu}} \ll \tau_0e^{1/\chi} \ .
\end{equation}
 Finally, the rate of the cavity growth is simply determined by
\begin{equation}
\dot{R}^{(0)}(t)=v_r^{(0)}(R^{(0)},t)  \ . \label{edge_velocity0}
\end{equation}

For the circular symmetry, the STZ equations
(\ref{eq:Dpl})-(\ref{Gamma}) reduce to
\begin{eqnarray}
\label{eq:Dpl0}
&\tau_0&\!\!\! {D^{pl}}^{(0)} = \epsilon_0 \Lambda^{(0)} \C C(s^{(0)})\left(\frac{s^{(0)}}{|s^{(0)}|}-m^{(0)}\right) \ , \\
\label{eq:m0} &\tau_0&\!\!\! \left(\frac{\partial m^{(0)}}{\partial
t}+v_r^{(0)}\frac{\partial m^{(0)}}{\partial r}\right)=\nonumber\\
&2&\!\!\!\frac{\tau_0 {D^{pl}}^{(0)}}{\epsilon_0 \Lambda^{(0)}
}- \Gamma^{(0)}(s^{(0)}, m^{(0)})m^{(0)}\frac{e^{-1/\chi^{(0)}}}{\Lambda^{(0)}} \ ,\\
\label{eq:Lambda0} &\tau_0&\!\!\! \left(\frac{\partial
\Lambda^{(0)}}{\partial t}+v_r^{(0)}\frac{\partial
\Lambda^{(0)}}{\partial r}\right) =\nonumber\\
&\Gamma&\!\!\!\!^{(0)}(s^{(0)}, m^{(0)})\left(e^{-1/\chi^{(0)}}- \Lambda^{(0)}\right) \ ,\\
\label{eq:chi0} &\tau_0&\!\!\! c_0 \left(\frac{\partial
\chi^{(0)}}{\partial t}+v_r^{(0)}\frac{\partial \chi^{(0)}}{\partial
r}\right) =\\
&\epsilon_0&\!\!\! \Lambda^{(0)} \Gamma^{(0)}(s^{(0)},
m^{(0)})\left[\chi_\infty\left(\tau_0{D^{pl}}^{(0)}\right)\! -\!
\chi^{(0)}\right] \ . \nonumber
\end{eqnarray}

Note that the $\chi$ and $D^{pl}$ equations contain a factor of the small STZ density $\epsilon_0 \Lambda^{(0)}$, which
implies they are much stiffer than the $m$ and $\Lambda$ equations.
Therefore, whenever the advection terms can be neglected this
separation of time scales \cite{07BLLP} allows us to replace the equations for
$m^{(0)}$ and $\Lambda^{(0)}$ by their stationary solutions
\begin{equation}
\label{m0_fix} m^{(0)}=\cases{ \frac{s^{(0)}}{|s^{(0)}|} &if $|
s^{(0)}|\le 1$\cr \frac{1}{ s^{(0)}} & if $| s^{(0)}| >1$}
\end{equation}
and
\begin{equation}
\Lambda^{(0)} = e^{-1/\chi^{(0)}} \ . \label{Lam0_fix}
\end{equation}
Note that Eq. (\ref{eq:m0}) has two stable fixed-point solutions given by
Eq. (\ref{m0_fix}), where we used Eq. (\ref{Lam0_fix}) and omitted the advection term. The transition between these two solutions corresponds to a transition between a jammed
and a plastically flowing state for a deviatoric stress below and above the yield stress respectively \cite{07BLanP}. Eq. (\ref{eq:Dpl0}) exhibits the corresponding solutions in terms of the plastic rate of deformation, zero and finite, below and above the yield stress respectively.

The unperturbed problem was studied in detail in Ref. \cite{07BLP}.
It was shown that for stresses $\sigma^\infty$ smaller than
a threshold value $\sigma^{th}\!\simeq\!5$ the cavity exhibits transient
dynamics in which its radius approaches a finite value in a finite
time. When this happens the material is jammed. On
the other hand, for $\sigma^\infty\!>\!\sigma^{th}$ the cavity  grows without bound, leading to a catastrophic failure of the material, accompanied by large scale plastic
deformations. We stress that to our knowledge this mode of failure by propagating a plastic solution
is new, apparently not related to other recently discovered failure fronts \cite{06GSW}.
One major goal of the present study is to analyze the
stability of the unbounded growth modes that result from this
cavitation. However, we are also interested in the range
$\sigma^\infty\!<\!\sigma^{th}$ where the unperturbed theory
predicts no catastrophic failure. In this range, a failure can still
occur if the cavity, prior to jamming, loses its perfect circular
symmetry in favor of relatively slender propagating ``fingers''. In
that case, stress localization near the tips of the propagating
``fingers'' can lead to failure via fracture. Such a scenario is typical of brittle fracture where the stress
localization due to the geometry of the defect drives crack
propagation that might lead to macroscopic failure.

\section{Linear Stability Analysis}
\label{LSA}

We derive here a set of equations for the linear
perturbations of the perfect circular symmetry where both inertia and
elastic compressibility effects are taken into account. In Appendix \ref{QS} we complement the analysis by considering the quasi-static and incompressible case. This case is mathematically more involved as it contains no explicit time evolution equation for the velocity and the pressure fields. By comparing the results of the two
formulations we test for consistency and obtain some degree of
confidence in the derivation and the numerical implementation of the
equations presented in this section.

\subsection{Equations of motion and kinematics}
\label{inertial}

The quantities involved in the problem are the tensors
\begin{equation}
\B s=\left(\begin{array}{cc}-s&\tau\\\tau&s\end{array}\right)\
,\quad
{\B
D}^{pl}=\left(\begin{array}{cc}-D^{pl}&D^{pl}_{r\theta}\\D^{pl}_{r\theta}&D^{pl}\end{array}\right)
\ ,
\end{equation}
as well as the pressure $p(\B r,t)$, the velocity $\B v(\B r,t)$ and
the location of the free boundary $R(\theta,t)$. We start by
expanding all these quantities as follows
\begin{eqnarray}
R(\theta, t)&=& R^{(0)}(t) + e^{i n \theta} R^{(1)}(t)\ ,\nonumber\\
s(r,\theta, t)&=&s^{(0)}(r,t) + e^{i n \theta} s^{(1)}(r,t)\ ,\nonumber\\
\tau (r,\theta, t)&=&i e^{i n \theta} \tau^{(1)}(r,t)\nonumber\\
p(r,\theta, t)&=&p^{(0)}(r,t) + e^{i n \theta} p^{(1)}(r,t)\ ,\nonumber\\
v_\theta(r,\theta, t)&=&i e^{i n \theta} v_\theta^{(1)}(r,t)\ ,\nonumber\\
v_r(r,\theta, t)&=&v_r^{(0)}(r,t) + e^{i n \theta} v_r^{(1)}(r,t)\ ,\nonumber\\
D^{pl}(r,\theta, t)&=& {D^{pl}}^{(0)}(r,t)+
e^{in\theta}{D^{pl}}^{(1)}(r,t)\ ,\nonumber\\
D^{pl}_{r\theta}(r,\theta, t) &=& i
e^{in\theta}{D^{pl}_{r\theta}}^{(1)}(r,t) \ .
\label{basic_perturbations}
\end{eqnarray}
Here all the quantities with the superscript $(1)$ are assumed to be
much smaller than their $(0)$ counterparts and $n$ is the discrete
azimuthal wave-number of the perturbations. The small perturbation hypothesis results in a formal linear decomposition in which each linear mode of wave-number $n$ is decoupled from all the other modes. When nonlinear contributions are non-negligible, all the modes become coupled and the formal linear decomposition is invalid.

We expand then the equations of motion
(\ref{EOM}) to first order to obtain
\begin{eqnarray}
\label{EOM1}
&&\rho \left(\frac{\partial v_r^{(1)}}{\partial t} +v_r^{(0)}
\frac{\partial v_r^{(1)}}{\partial r}+v_r^{(1)} \frac{\partial
v_r^{(0)}}{\partial r} \right)= \nonumber\\
&&-\frac{n\tau^{(1)}}{r} -\frac{1}{r^2} \frac{\partial }{\partial r}
\left ( r^2 s^{(1)} \right) - \frac{\partial p^{(1)}}{\partial r} \
, \\\label{EOM2} &&\rho \left(\frac{\partial
v_\theta^{(1)}}{\partial t} +v_r^{(0)}
\frac{\partial v_\theta^{(1)}}{\partial r}+
\frac{v_r^{(0)} v_\theta^{(1)}}{r}\right)=\nonumber\\
&&\frac{\partial \tau^{(1)}}{\partial r} + \frac{n s^{(1)}}{r}
 - \frac{n p^{(1)}}{r}
 +\frac{2 \tau^{(1)}}{r} \ .
\end{eqnarray}
We proceed by expanding Eqs. (\ref{el_pl}) to first order, which after a simple manipulation yields
\begin{eqnarray}
\label{first_eq}&&\frac{\partial v^{(1)}_r}{\partial r}+\frac{-n v^{(1)}_\theta +
v^{(1)}_r}{r}=\\
&& -\frac{1}{K}\left(\frac{\partial p^{(1)}}{\partial t}+v_r^{(0)}
\frac{\partial p^{(1)}}{\partial r}+v_r^{(1)}
\frac{\partial p^{(0)}}{\partial r}\right)\ ,\nonumber\\
&&\frac{-n v^{(1)}_\theta + v^{(1)}_r}{r}-\frac{\partial
v^{(1)}_r}{\partial r} =\\
&&\frac{1}{ \mu} \left[ \frac{\partial s^{(1)}}{\partial t} +
v_r^{(0)} \frac{\partial s^{(1)}}{\partial r} + v^{(1)}_r
\frac{\partial s^{(0)}}{\partial r}\right] + 2{D^{pl}}^{(1)} \ ,
\nonumber\\
&&\frac{1}{2} \left[\frac {\partial v^{(1)}_\theta}{\partial r} +
\frac{n v^{(1)}_r - v^{(1)}_\theta}{r} \right] = \\
&&\frac{1}{2 \mu} \left[ \frac{\partial \tau^{(1)} }{\partial t} +
v^{(0)}_r \frac{\partial \tau^{(1)}}{\partial r} -\frac{2 s^{(0)} v_\theta^{(1)}}{r}\right]
 + {D^{pl}_{r \theta}}^{(1)}\nonumber \ . \label{last_eq}
\end{eqnarray}

At this point we derive an
evolution equation for the dimensionless amplitude of the shape
perturbation $R^{(1)}/R^{(0)}$. To that aim we note that
\begin{equation}
\dot{R}=v_r(R)+{\cal O}\left[\left(\frac{R^{(1)}}{R^{(0)}} \right)^2
\right] \ .
\end{equation}
Expanding this relation using Eqs. (\ref{basic_perturbations}), we
obtain to zeroth order Eq. (\ref{edge_velocity0}) and to first order
\begin{equation}
\dot{R}^{(1)}(t)=v_r^{(1)}(R^{(0)})+R^{(1)} \frac{\partial
v_r^{(0)}(R^{(0)})}{\partial r} \ . \label{edge_velocity1}
\end{equation}
Therefore, we obtain
\begin{eqnarray}
\label{smallness}
&&\frac{d}{dt}\left(\frac{R^{(1)}}{R^{(0)}}
\right)=\\
&&\frac{R^{(1)}}{R^{(0)}}\!\left[\!\frac{v_r^{(1)}(R^{(0)})}{R^{(1)}}\!+\!\frac{\partial
v_r^{(0)}(R^{(0)})}{\partial r}\!-\!
\frac{v_r^{(0)}(R^{(0)})}{R^{(0)}}\!\right] \ .\nonumber
\end{eqnarray}
This is an important equation since a linear instability manifests itself
as a significant increase in $R^{(1)}/R^{(0)}$ such that
nonlinear terms become non-negligible. Note that the two last terms
in the square brackets are always negative, therefore an instability
can occur only if the first term in the square brackets is positive
with absolute value larger than the sum of the two negative terms.
Moreover, recall that the problem is non-stationary, implying that
all the zeroth order quantities depend on time.

In order to derive the boundary conditions for the components of the
stress tensor field we expand to linear order the normal unit vector
$\B n$ (not to be confused with the discrete wave-number $n$) and
tangential unit vector $\B t$ at the free boundary, obtaining
\begin{equation}
\label{unit_n} \B n = \left( 1, -i \frac{R^{(1)}}{R^{(0)}}ne^{in
\theta} \right)\ ,\quad \label{unit_t} \B t = \left(i
\frac{R^{(1)}}{R^{(0)}}ne^{in \theta} , 1 \right) \ .
\end{equation}
Eqs. (\ref{stressBC}), expanded to first order, translate to
\begin{eqnarray}
\label{spb}
s^{(1)}(R^{(0)})\!\!&+&\!\!p^{(1)}(R^{(0)})=\nonumber\\
\!\!&-&\!\!R^{(1)}\left[\frac{\partial
s^{(0)}(R^{(0)})}{\partial r}+ \frac{\partial p^{(0)}(R^{(0)})}{\partial r}\right]\ , \\
\label{taub} \tau^1(R^{(0)})\!\!&=&\!\! n
\left[s^{(0)}(R^{(0)})-p^{(0)}(R^{(0)})\right]
\frac{R^{(1)}}{R^{(0)}}.
\end{eqnarray}
In addition, all the first
order fields decay as $r\!\to\!\infty$. The initial conditions are determined by the
perturbation scheme that is being studied.

To avoid dealing with an infinite and time-dependent domain we applied
the following time-dependent coordinate transformation
\begin{equation}
\xi=R(t)/r \ . \label{trans}
\end{equation}
This transformation allows us to integrate the equations in the
time-independent finite domain $\xi\! \in\! [0,1]$, with the price
of introducing new terms in the equations. Controlling the equations at
small distances required the introduction of an artificial viscosity
on the right-hand-side (RHS) of Eq. (\ref{eqmot1}). The term
introduced is $\rho\eta\! \nabla^2\! \B v$, with $\eta$ chosen of
the order of the square of space discretization over the time
discretization. This introduces zeroth order contributions on the
RHS of Eq. (\ref{EOM0}) and first order contributions on the RHS of
Eqs. (\ref{EOM1})-(\ref{EOM2}).

\subsection{Linear perturbation analysis of the STZ equations}
\label{perturbSTZ}

The only missing piece in our formulation is the perturbation of
the tensorial STZ equations. In addition to the fields considered up
to now, the analysis of the STZ equations includes also the
internal state fields
\begin{equation}
{\bf m}=\left(\begin{array}{cc}-m&m_{r\theta}\\m_{r\theta}&m
\end{array}\right),\quad\Lambda\quad\hbox{and}\quad\chi \ .
\end{equation}
Therefore, in addition to Eqs. (\ref{basic_perturbations}) we have
\begin{eqnarray}
m(r,\theta, t)&=& m^{(0)}(r,t)+
e^{in\theta}m^{(1)}(r,t)\ ,\nonumber\\
m_{r\theta}(r,\theta, t)&=&i e^{i n \theta} m_{r\theta}^{(1)}(r,t)\ ,\nonumber\\
\Lambda(r,\theta, t)&=& \Lambda^{(0)}(r,t)+e^{in\theta}\Lambda^{(1)}(r,t)\ ,\nonumber\\
\chi(r,\theta,t) &=& \chi^{(0)}(r,t) + e^{in\theta}\chi^{(1)}(r,t)
\ . \label{STZ_perturbations}
\end{eqnarray}

We then expand systematically Eqs. (\ref{eq:Dpl})-(\ref{C_s}).
First, we have
\begin{eqnarray}
\label{sbar_exp}
\bar{s}&=&\sqrt{\frac{2(s^{(0)}+e^{in\theta}s^{(1)})^2 +
2(\tau^{(1)} e^{in\theta})^2}{2}} \\
&\simeq& |s^{(0)} +
e^{in\theta}s^{(1)}|=|s^{(0)}|+e^{in\theta}s^{(1)}{\rm
sgn}\left(s^{(0)}\right) \ .\nonumber
\end{eqnarray}
Accordingly we expand $\C C(\bar s)$ (assuming $s^{(0)}>0$) in the
form
\begin{equation}
\C C(\bar s) = \C C(s^{(0)} + e^{in\theta}s^{(1)}) \simeq \C
C(s^{(0)}) + \frac{d\C
C}{ds}\left(s^{(0)}\right)e^{in\theta}s^{(1)},
\end{equation}
where
\begin{equation}
\frac{d\C C}{ds}\left(s^{(0)}\right) =
\frac{\zeta^{\zeta+1}}{\zeta!}\int_0^{|s^{(0)}|}s_\alpha^\zeta
exp(-\zeta s_\alpha)ds_\alpha \ .
\end{equation}
Substituting the last three equations into (\ref{eq:Dpl}) and
expanding to first order, we obtain
\begin{widetext}
\begin{eqnarray}
\label{Dpl1} \tau_0{D^{pl}}^{(1)} &=&
\epsilon_0\Lambda^{(0)}\left[\left(\frac{\Lambda^{(1)}}{\Lambda^{(0)}}\C
C\left(s^{(0)}\right) + s^{(1)}\frac{d\C
C\left(s^{(0)}\right)}{ds}\right)\left(
{\rm
sgn}\left(s^{(0)}\right)-m^{(0)}\right)
- \C C\left(s^{(0)}\right)m^{(1)}\right] \ , \\
\label{Dploff1} \tau_0 {D^{pl}_{r\theta}}^{(1)} &=& \epsilon_0
\Lambda^{(0)}\C
C\left(s^{(0)}\right)\left(\frac{\tau^{(1)}}{|s^{(0)}|}-m_{r\theta}^{(1)}\right)
\ .
\end{eqnarray}
\end{widetext}
We then expand $\Gamma$ in the form
\begin{eqnarray}
\Gamma \!&=&\! \Gamma^{(0)} +
e^{in\theta}\Gamma^{(1)}\quad\hbox{with}\quad \Gamma^{(0)} =
\frac{2 \tau_0
s^{(0)}{D^{pl}}^{(0)}}{\epsilon_0 \Lambda^{(0)}} \ ,\nonumber\\
\Gamma^{(1)}\! &=&\! \frac{2 \tau_0 }{\epsilon_0 \Lambda^{(0)}}
\left[s^{(0)}{D^{pl}}^{(1)}\! +\! s^{(1)}{D^{pl}}^{(0)}
\!-\!\frac{s^{(0)}{D^{pl}}^{(0)}\Lambda^{(1)}}{\Lambda^{(0)}}\right]
\
.\nonumber\\
\end{eqnarray}
Eq. (\ref{eq:m}) is now used to obtain
\begin{eqnarray}
\label{m1} &&\tau_0\left(\frac{\partial m^{(1)}}{\partial t}
+v_r^{(0)} \frac{\partial m^{(1)}}{\partial r}+v_r^{(1)}
\frac{\partial
m^{(0)}}{\partial r} \right) = \nonumber\\
&&\frac{2\tau_0}{\epsilon_0 \Lambda^{(0)}}\left( {D^{pl}}^{(1)} -
{D^{pl}}^{(0)}\frac{\Lambda^{(1)}}{\Lambda^{(0)}}\right)-\frac{e^{-1/\chi^{(0)}}}{\Lambda^{(0)}}\times\\
&&\left[\Gamma^{(0)}
m^{(1)} +\Gamma^{(1)}
m^{(0)}+ \Gamma^{(0)}
m^{(0)}\left(\frac{\chi^{(1)}}{\left[\chi^{(0)}\right]^2}-
\frac{\Lambda^{(1)}}{\Lambda^{(0)}}\right) \right] \ ,\nonumber
\end{eqnarray}
and
\begin{eqnarray}
\label{moff1} &&\tau_0\left(\frac{\partial
m_{r\theta}^{(1)}}{\partial t} +v_r^{(0)} \frac{\partial
m_{r\theta}^{(1)}}{\partial r}-\frac{m^{(0)} v_\theta^{(1)}}{r} \right) =\nonumber\\
&& \frac{2\tau_0
{D^{pl}_{r\theta}}^{(1)}}{\epsilon_0 \Lambda^{(0)}} -\Gamma^{(0)}
m_{r\theta}^{(1)}\frac{e^{-1/\chi^{(0)}}}{\Lambda^{(0)}} \ .
\end{eqnarray}
Using Eq. (\ref{eq:Lambda}) we obtain
\begin{eqnarray}
\label{Lambda1} &\tau_0&\!\!\!\!\! \left(\frac{\partial
\Lambda^{(1)}}{\partial t} +v_r^{(0)} \frac{\partial
\Lambda^{(1)}}{\partial r}+v_r^{(1)} \frac{\partial
\Lambda^{(0)}}{\partial r} \right)= \\
&\Gamma^{(0)}&\!\!\!\left(e^{-1/\chi^{(0)}}\frac{\chi^{(1)}}{\left[\chi^{(0)}\right]^2}\!-\!\Lambda^{(1)}
\right)\!+\!\Gamma^{(1)}\left(e^{-1/\chi^{(0)}}\!-\!\Lambda^{(0)}
\right) \ . \nonumber
\end{eqnarray}
Expanding $\bar{D}^{pl}$, similarly to Eq. (\ref{sbar_exp}), we
obtain
\begin{equation}
\bar{D}^{pl} \simeq
|{D^{pl}}^{(0)}|+e^{in\theta}{D^{pl}}^{(1)}{\rm
sgn}\left({D^{pl}}^{(0)}\right) \ .
\end{equation}
Accordingly we expand $\chi_\infty\left(\tau_0 \bar{D}^{pl}
\right)$ (with ${D^{pl}}^{(0)}\!>\!0$) in the form
\begin{eqnarray}
&&\chi_\infty\left(\tau_0 \bar{D}^{pl} \right) =
\chi_\infty\left(\tau_0 {D^{pl}}^{(0)} + e^{in\theta}\tau_0
{D^{pl}}^{(1)}\right) \\
&&= \chi_\infty\left(\tau_0 {D^{pl}}^{(0)}\right) +
\frac{d\chi_\infty}{d\bar{D}^{pl}}\left(\tau_0
{D^{pl}}^{(0)}\right)e^{in\theta} {D^{pl}}^{(1)} \
.\nonumber
\end{eqnarray}
Then, using Eq. (\ref{eq:chi}) we obtain
\begin{widetext}
\begin{eqnarray}
\label{chi1} &&\tau_0 c_0 \left(\frac{\partial
\chi^{(1)}}{\partial t} +v_r^{(0)} \frac{\partial
\chi^{(1)}}{\partial r}+v_r^{(1)} \frac{\partial
\chi^{(0)}}{\partial r} \right)= \epsilon_0\left(\Lambda^{(0)}\Gamma^{(1)}+\Gamma^{(0)}\Lambda^{(1)}\right)
\left(\chi_\infty\left(\tau_0{D^{pl}}^{(0)}\right)-\chi^{(0)}\right)+\nonumber\\
&&\epsilon_0\Lambda^{(0)}\Gamma^{(0)}\left(\frac{d\chi_\infty}{d\bar{D}^{pl}}\left(\tau_0
{D^{pl}}^{(0)}\right){D^{pl}}^{(1)}-\chi^{(1)}\right)\ .
\end{eqnarray}
\end{widetext}
Thus, Eqs. (\ref{Dpl1})-(\ref{Dploff1}), (\ref{m1})-(\ref{moff1}),
(\ref{Lambda1}) and (\ref{chi1}) constitute our equations for the
dynamics of the first order STZ quantities.

These equations already reveal some interesting features. First note
that the coupling between ${D^{pl}}^{(1)}$ (which is the quantity
that is expected to be of major importance in determining
$v_r^{(1)}$ in Eq. (\ref{smallness}) through Eqs.
(\ref{first_eq})-(\ref{last_eq})) and $\chi^{(1)}$, $m^{(1)}$
depends on $\C C\left(s^{(0)}\right)$. This means that the strength
of the coupling depends on $\zeta$. Similarly, the coupling between
${D^{pl}}^{(1)}$ and $s^{(1)}$ depends on $d\C
C\left(s^{(0)}\right)/ds$ which is also a function of $\zeta$. These observations demonstrate the importance of the precise form of the function $\C C(s)$. This issue is further discussed in Sec. \ref{rate_function}. Finally, note that whenever
the advection terms can be neglected, the known separation of time
scales \cite{07BLLP} allows us to use Eqs. (\ref{m0_fix})-(\ref{Lam0_fix}) and to
replace the equations for $m^{(1)}$, $m_{r\theta}^{(1)}$ and
$\Lambda^{(1)}$ by their stationary solutions
\begin{equation}
\label{m1_fix} m^{(1)}=\cases{ 0 &if $s^{(0)}\le 1$\cr
-\frac{s^{(1)}}{\left[s^{(0)}\right]^2} & if $ s^{(0)} >1$} \ ,
\end{equation}
\begin{equation}
\label{m_rth_fix} m_{r\theta}^{(1)}=\cases{
\frac{\tau^{(1)}}{s^{(0)}} &if $s^{(0)}\le 1$\cr
\frac{\tau^{(1)}}{\left[s^{(0)}\right]^2} & if $s^{(0)} >1$}
\end{equation}
and
\begin{equation}
\Lambda^{(1)} =
\frac{\chi^{(1)}}{\left[\chi^{(0)}\right]^2}e^{-1/\chi^{(0)}} \ .
\label{Lam1_fix}
\end{equation}

In the next section we summarize the results of our analysis of the
equations derived in Sec. \ref{zeroth}, \ref{inertial} and
\ref{perturbSTZ}.

\section{Results}
\label{results}

We are now ready to present and discuss the results of the stability analysis of
the expanding circular cavity. The full set of equations was solved numerically as
discussed above. Time and length are measured in units of $\tau_0$
and $R^{(0)}(t\!=\!0)$ respectively. $\Lambda$ and $m$ are set
initially to their respective fixed-points. The material-specific
parameters used are $\epsilon_0\!=\!1$, $c_0\!=\!1$,
$\mu/s_y\!=\!50$, $K/s_y\!=\!100$, $\rho=1$, $\chi^{(0)}\!=\!0.11$,
$\chi_\infty\!=\!0.13$ and $\zeta\!=\!7$, unless otherwise stated. In Subsec.
\ref{pert_shape} we study perturbations of the shape of the cavity and of the effective temperature $\chi$.
In Subsec. \ref{strain_rate} we study the effect of the rate dependence of $\chi_\infty$ on the stability analysis and in Subsec. \ref{rate_function}
we analyze the effect of the stress-dependent rate function $\C C(s)$.

\subsection{Perturbing the shape and $\chi$}
\label{pert_shape}

\begin{figure}
\centering \epsfig{width=.47\textwidth,file=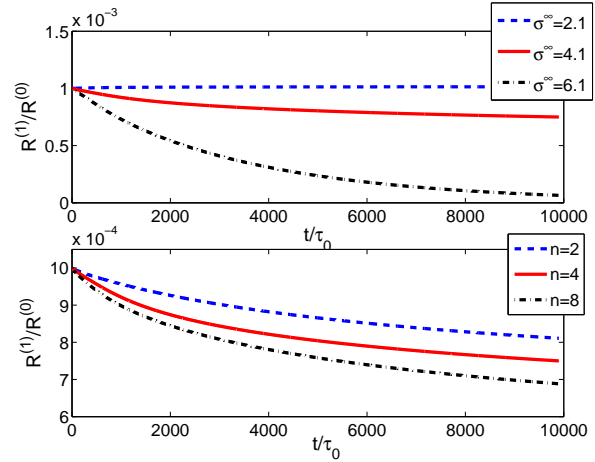}
\caption{(Color online) Upper panel: The ratio $R^{(1)}/R^{(0)}$ as
a function of time for $n\!=\!4$ and $\sigma^\infty\!=\!2.1,4.1$ and 6.1. Note
that the last value is above the cavitation threshold. Lower panel:
The ratio $R^{(1)}/R^{(0)}$ as a function of time for $\sigma^\infty\!=\!4.1$ and $n\!=\!2,4$
and 8.}\label{pertR}
\end{figure}

Studying the linear stability of the expanding cavity can be done by selecting
which fields are perturbed and which are left alone. In practice
each of the fields involved in
the problem may experience simultaneous fluctuations, including the
radius of the cavity itself. Therefore, one of our tasks is to determine which of the possible
perturbation leads to a linear instability. To start, we
perturb the radius of the expanding cavity at $t\!=\!0$ while all the other fields
are left alone. In Fig. \ref{pertR} we
show the ratio $R^{(1)}/R^{(0)}$ as a function of time for various
loading levels $\sigma^\infty$ (both below and above the cavitation
threshold) and wave-numbers $n$. The initial amplitude of the
perturbation was set to $R^{(1)}/R^{(0)}\!=\!10^{-3}$. The observation is that
the ratio $R^{(1)}/R^{(0)}$ does not grow in time in all the considered cases were the radius was perturbed, implying that here the circular
cavity is stable against shape perturbations. Note
that $R^{(1)}/R^{(0)}$ decays faster for larger n and for larger
$\sigma^\infty$. Also note that for $\sigma^\infty\!=\!6.1$. i.e.
for unbounded zeroth order expansion, the ratio $R^{(1)}/R^{(0)}$
decays to zero while below the cavitation threshold this ratio
approaches a finite value. The latter observations means that when the material
approaches jamming (with $R^{(0)}$ attaining a finite value in a
finite time) the perturbation have not yet disappeared
entirely.

\begin{figure}
\centering \epsfig{width=.47\textwidth,file=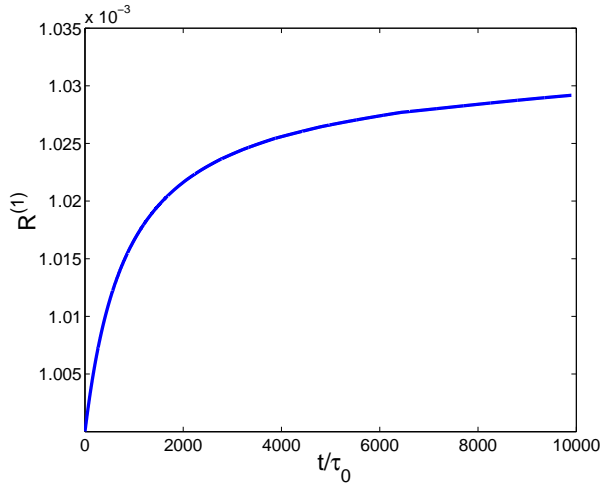}
\caption{(Color online) $R^{(1)}$ as a function of time for
$\sigma^\infty\!=\!2.1$ and $n\!=\!4$.}\label{onlyR1}
\end{figure}

We stress at this point the non-stationary nature of
the problem in which $R^{(0)}(t)$ is an increasing function of time.
Thus, even if the absolute magnitude of the amplitude of
the shape perturbation $R^{(1)}(t)$ increases with time, an
instability is not automatically implied; $R^{(1)}(t)$ should increase
sufficiently faster than $R^{(0)}(t)$ in order to imply an
instability. To exemplify this feature of the problem, we present in
Fig. \ref{onlyR1} $R^{(1)}(t)$ for $\sigma^\infty\!=\!2.1$ and $n\!=\!4$. It is observed that even though
$R^{(1)}$ increases, the  smallness parameter
$R^{(1)}/R^{(0)}$ decreases, see Fig. \ref{pertR}. Note also that
$R^{(1)}$ does not increase exponentially as expected in stationary
linear stability analysis, but rather tends to asymptote to a
constant.

Next we have tested the stability of the expanding cavity against initial
perturbations in the velocity field or in the stress field. The results were quantitatively similar to those for
the shape perturbations summarized in Figs. \ref{pertR} and
\ref{onlyR1}, all implying linear stability.

In light of these
results, we concentrated then on the effect of perturbations in the STZ
internal state fields. Since the dynamics of the tensor $\B m$ are
mainly determined by the deviatoric stress field $s$, we focus on
fluctuations in the effective disorder temperature $\chi$. This may be the
most liable field to cause an instability. Indeed,
in Ref. \cite{07MLC} it was shown that $\chi$ perturbations control strain localization in
a shear banding instability. Qualitatively, an instability in the
form of growing ``fingers'' involves strain localization as well;
plastic deformations are localized near the leading edges of the
propagating ``fingers''. In Ref. \cite{07MLC}, based on the data of Ref.
\cite{07SKLF}, it was suggested that the typical spatial fluctuations in
$\chi$ have an amplitude reaching about $30\%$ of the homogeneous
background $\chi$. Obviously we cannot treat such large
perturbations in a linear analysis and must limit ourselves to
smaller perturbations.
\begin{figure}
\centering \epsfig{width=.47\textwidth,file=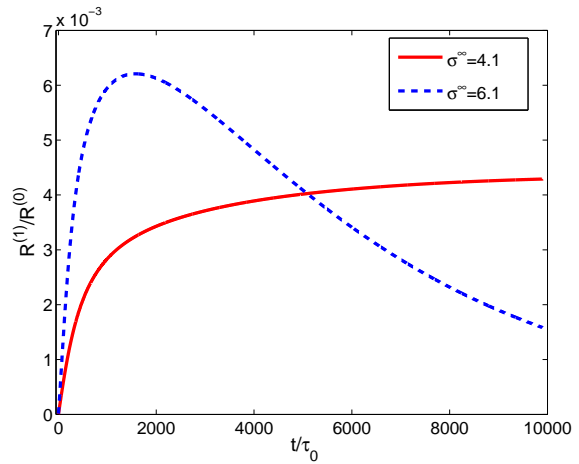}
\caption{(Color online) The ratio $R^{(1)}/R^{(0)}$ as a function of
time for a perturbation of size $\chi^{(1)}/\chi^{(0)}\!=\!0.03$ and
$n\!=\!4$, introduced at time $t\!=\!0$. The solid line corresponds
to $\sigma^\infty\!=\!4.1$ (below the cavitation threshold) and the
dashed line corresponds to $\sigma^\infty\!=\!6.1$ (above the
cavitation threshold).}\label{pertChi}
\end{figure}
In Fig. \ref{pertChi} we show the ratio $R^{(1)}/R^{(0)}$ as a function of time for a perturbation of
size $\chi^{(1)}/\chi^{(0)}\!=\!0.03$, introduced at time $t\!=\!0$.
The wave-number was set to $n\!=\!4$ and $\sigma^\infty$ was set
both below and above the cavitation threshold. First, note that for
both loading conditions $R^{(1)}/R^{(0)}$ increases on a short time
scale of about $1000\tau_0$, a qualitatively different behavior
compared to the system's response to shape perturbations. Second,
note the qualitatively different response below and above the
cavitation threshold. In the former case, $R^{(1)}/R^{(0)}$
increases monotonically, approaching a constant value when $R^{(0)}$
attains a finite value (i.e. jamming). In the latter case,
$R^{(1)}/R^{(0)}$ increases more rapidly initially, reaches a
maximum and then decays to 0 in the large $t$ limit. Therefore, in
spite of the initial growth of $R^{(1)}/R^{(0)}$, for this magnitude
of $\chi$ perturbations, the expanding circular cavity is linearly
stable; below the cavitation threshold the relative magnitude of the
deviation from a perfect circular symmetry $R^{(1)}/R^{(0)}$ tends
to a finite constant, i.e. a shape perturbation is ``locked in'' the
material, while above the threshold the cavity retains its perfect
circular symmetry in the large $t$ limit. Nevertheless, in light of
the significant short time increase in $R^{(1)}/R^{(0)}$ (here up to
$0.6\%$), we increased the initial ($t\!=\!0$) $\chi$ perturbation
to the range $\chi^{(1)}/\chi^{(0)}\!=\!0.05-0.06$, in addition to shape
perturbations of a typical size of $R^{(1)}/R^{(0)}\!=\!0.02-0.03$.
In these cases $R^{(1)}/R^{(0)}$ grows above $5\%$; even more
importantly, the field $\chi^{(1)}(\B r, \theta)$ (as well as other
fields in the problem) becomes larger than $0.1\chi^{(0)}(\B r,
\theta)$ near the boundary of the cavity, {\em invalidating} the
small perturbation hypothesis behind the perturbative expansion and
signaling a linear instability. Naturally, this breakdown of the
linearity condition  takes place firstly near a peak of the ratio
$R^{(1)}/R^{(0)}$, similar to the one observed in Fig.
\ref{pertChi}.

We thus propose that sufficiently large
perturbations in the shape of the cavity and the effective disorder
temperature $\chi$, but still of formal linear order, may lead
to an instability. This dependence on the magnitude of the
perturbations in a linear analysis is a result of the
non-stationarity of the growth. Another manifestation of the
non-stationarity is that even in cases where we
detect an instability, it was not of the usual simple exponential
type where an eigenvalue changes sign as a function of some
parameter (or group of parameters). Combined with the evidence for
the existence of large fluctuations in $\chi$ \cite{07SKLF, 07MLC}, the present results
indicate that it will be worthwhile to study the problem by direct boundary tracking
techniques where the magnitude of the perturbation is not limited.

We conclude that the issue of the stability of
the expanding cavity can be subtle. Sufficiently small
perturbations are stable, though there is a qualitative difference
in the response to perturbations in the effective disorder
temperature $\chi$, where the ratio $R^{(1)}/R^{(0)}$ increases (at
least temporarily), and other perturbations, where $R^{(1)}/R^{(0)}$
decays. We have found that for large enough $\chi$ perturbations
combined with initial shape perturbations, but still within the
formal linear regime, the growth of $R^{(1)}/R^{(0)}$ takes the system
beyond the linear regime, making nonlinear effects non-negligible
and signaling an instability. This observation is further supported by the
existence of large $\chi$ fluctuations discussed in \cite{07SKLF, 07MLC}.
Note that none of these conclusions depend significantly on variations in
$\epsilon_0$ and $c_0$. Moreover, perturbing the expanding
cavity at times different than  $t\!=\!0$ or introducing a pressure
inside the cavity instead of a tension at infinity did not change any of the results.

\subsection{The effect of the rate dependence of $\chi_\infty$}
\label{strain_rate}

\begin{figure}
\centering \epsfig{width=.47\textwidth,file=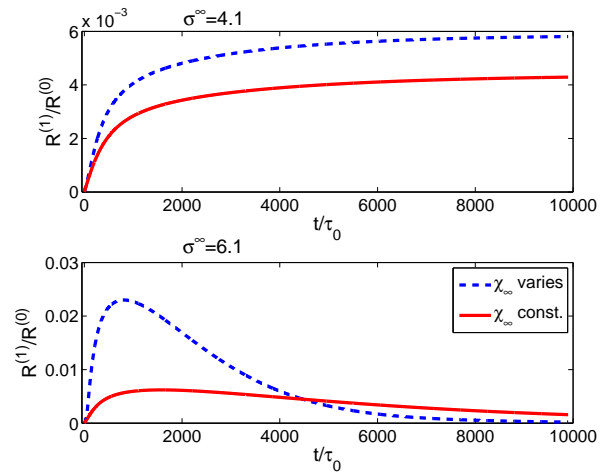}
\caption{(Color online) Upper panel: $R^{(1)}/R^{(0)}$ as a function
of time for $\sigma^\infty\!=\!4.1$ (below the cavitation
threshold). The solid line corresponds to a constant $\chi_\infty$
and the dashed line corresponds to the plastic rate of deformation
dependent $\chi_\infty(\tau_0 \bar{D}^{pl})$ of Fig. \ref{HL}. The
initial perturbation has $\chi^{(1)}/\chi^{(0)}\!=\!0.03$ and
$n\!=\!4$. Lower panel: The same with $\sigma^\infty\!=\!6.1$ (above
the cavitation threshold).}\label{StrainRate}
\end{figure}

The analysis of Sec. \ref{pert_shape} indicates the existence of a linear
instability as a result of varying the magnitude of the
perturbations, mainly in $\chi$, and not as a result of varying
material parameters. Here, and in Sec. \ref{rate_function}, we aim
at studying the effect of material-specific properties on the stability
of the expanding cavity. Up to now we considered $\chi_\infty$ as a constant parameter. However, as discussed in detail in Sec.
\ref{STZ}, the plastic rate of deformation near the free boundary
can reach values in the range where changes in $\chi_\infty$ were
observed. Therefore, we repeated the calculations using the function
$\chi_\infty(\tau_0 \bar{D}^{pl})$ plotted in Fig. \ref{HL}. In Fig.
\ref{StrainRate} we compare $R^{(1)}/R^{(0)}$ as a function of time
with and without a plastic rate of deformation dependence of
$\chi_\infty$, both above and below the cavitation threshold. The
initial perturbation has $\chi^{(1)}/\chi^{(0)}\!=\!0.03$ and
$n\!=\!4$.
Both below and above the cavitation threshold the plastic rate
of deformation dependent $\chi_\infty(\tau_0 \bar{D}^{pl})$ induces
a stronger growth of $R^{(1)}/R^{(0)}$, though the effect is
much more significant above the threshold. This is understood as
significantly higher rate of deformation is developed above the
cavitation threshold, where unbounded growth takes place \cite{07BLP},
compared to below the threshold where the rate of deformation
vanishes at a finite time. We note that the dependence of $\chi_\infty$
on $ \bar{D}^{pl}$ affects both the zeroth and
first order solutions such that $R^{(0)}$ and $R^{(1)}$ increase.
Our results show that $R^{(1)}$ is more sensitive to this effect
than $R^{(0)}$, resulting in a tendency to lose stability at yet
smaller perturbations. We conclude that the tendency of $\chi_\infty$
to increase with the rate of deformation plays an important role in
the stability of the expanding cavity and might be crucial for other
strain localization phenomena as the shear banding instability
\cite{07MLC}. Moreover, this material-specific dependence of $\chi_\infty$, that was absent in previous
formulations of STZ theory, might distinguish between materials that
experience catastrophic failure and those that do not, and between
materials that fail through a cavitation instability \cite{07BLP} and
those who fail via the propagation of ``fingers'' that may evolve
into cracks. This new aspect of the theory certainly deserves more
attention in future work. We note in passing that recently an alternative equation to Eq. (\ref{eq:chi}) for the time evolution of the effective temperature $\chi$ was proposed in light of some available experimental and simulational data \cite{08Bouch}. Preliminary analysis of the new equation in relation to the stability analysis performed in this paper indicates that the circular cavity {\em does} become linearly unstable \cite{unpublished}. A more systematic study of this effect may be a promising line of future investigation.

\subsection{The effect of changing the stress-dependent rate function $\C C(s)$}
\label{rate_function}

\begin{figure}
\centering \epsfig{width=.47\textwidth,file=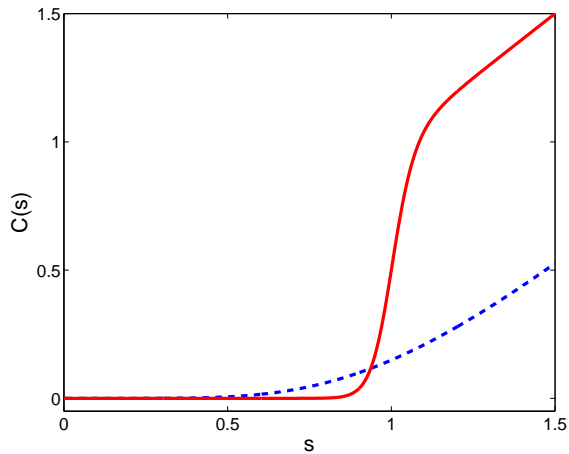}
\caption{(Color online) The function $\C C(s)$ of Eq. (\ref{C_s})
with $\zeta\!=\!7$ (dashed line) and of Eq. (\ref{C_s1}) with
$\lambda\!=\!30$ (solid line).}\label{changingC}
\end{figure}

Here we further study the possible effects of details of the
constitutive behavior on the macroscopic behavior of the expanding
cavity. In this subsection we focus on the material function $\C C(s)$.
This phenomenological function, as discussed in Sec. \ref{STZ},
describes the stress-dependent STZ transition rates. It is expected
to be symmetric and to vanish smoothly at $s\!=\!0$ in
athermal conditions \cite{07BLanP}. The plastic rate of deformation
for $s\!>\!1$ can be measured in a steady state stress-controlled
simple shear experiment. For such a configuration the deviatoric
stress tensor is diagonal and the stable fixed-points of Eqs.
(\ref{eq:m})-(\ref{eq:chi}) imply that the steady state plastic rate
of deformation of Eq. (\ref{eq:Dpl}) reads
\begin{equation}
\label{steadyDpl} \tau_0 D^{pl} = \epsilon_0 e^{-1/\chi_\infty} \C
C(s)\left(1-\frac{1}{s}\right) \ .
\end{equation}
Therefore, if the steady state relation $\chi_\infty(s)$ is known, $\C C(s)$ can be determined from measuring the steady state value of $D^{pl}$ for various $s\!>\!1$, see for example \cite{07HL}. The idea then is to interpolate
the $s\!\to\!0^+$ behavior to the $s\!>\!1$ behavior with a single
parameter that controls the amount of sub-yield deformation in the
intermediate range. In fact, a procedure to measure $\C C(s)$ at
intermediate stresses was proposed in Ref. \cite{07BL}. Up to now
we used the one-parameter family of functions $\C F(s;\zeta)$
of Eq. (\ref{C_s}), where $\zeta$ controls the sub-yield
deformation.

We now aim at studying the effect of choosing another function
${\cal C}(s)$. Here we specialize for ${\cal
C}(\bar{s})=\C G(\bar{s}; \lambda)$, with
\begin{equation}
\label{C_s1} \C G(\bar{s};\lambda)\equiv
\frac{|\bar{s}|^{1+\lambda}}{1+|\bar{s}|^\lambda} \ .
\end{equation}
In Fig. \ref{changingC} we show $\C C(s)$ according to the previous choice
of Eq. (\ref{C_s}) with $\zeta\!=\!7$ and also $\C C(s)$ according
to the present choice of Eq. (\ref{C_s1}) with $\lambda\!=\!30$. The
different behaviors of $\C C(s)$ and $d\C C(s)/ds$ near $s\!=\!1$
might affect differently $R^{(0)}$ and $R^{(1)}$, thus influencing
the stability of the expanding cavity.

\begin{figure}
\centering \epsfig{width=.47\textwidth,file=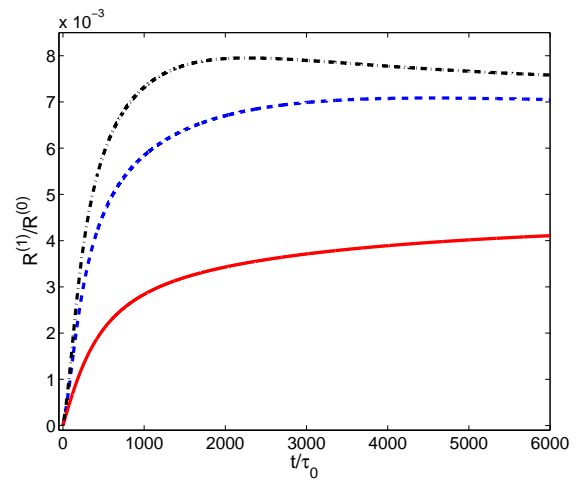}
\caption{(Color online) $R^{(1)}/R^{(0)}$ as a function of time for
an effective temperature perturbation with
$\chi^{(1)}/\chi^{(0)}\!=\!0.03$ and $n\!=\!4$ for $\sigma^\infty\!=\!4.1$. The solid line
corresponds to $\C C(\bar s)$ of Eq. (\ref{C_s}) with $\zeta\!=\!7$
and the dotted line corresponds to $\C C(\bar s)$ of Eq.
(\ref{C_s1}) with $\lambda\!=\!30$, see Fig. \ref{changingC}. In
both cases a constant $\chi_\infty$ was used. The dotted-dashed line
corresponds to $\C C(\bar s)$ of Eq. (\ref{C_s1}) with
$\lambda\!=\!30$ and the rate dependent
$\chi_\infty(\tau_0\bar{D}^{pl})$ presented in Fig.
\ref{HL}.}\label{changingC1}
\end{figure}

In Fig. \ref{changingC1} we compare $R^{(1)}/R^{(0)}$ as a function
of time for $\C C(s)$ of Eq. (\ref{C_s}) (previous choice) with
$\zeta\!=\!7$ and $\C C(s)$ of Eq. (\ref{C_s1}) (present choice)
with $\lambda\!=\!30$, both for a constant $\chi_\infty$. An effective temperature perturbation with
$\chi^{(1)}/\chi^{(0)}\!=\!0.03$ and $n\!=\!4$ was introduced at
$t\!=\!0$ for $\sigma^\infty\!=\!4.1$. We observe that $R^{(1)}/R^{(0)}$ grows faster for the
present choice compared to the previous one. For the sake of
illustration we added the result of a calculation with the plastic
rate of deformation dependent $\chi_\infty$ as discussed in Sec.
\ref{strain_rate}. As expected, the effect is magnified. We conclude
that the material-specific function of the stress dependence of the
STZ transition rates $\C C(s)$ can affect the stability of the
expanding cavity, possibly making it unstable for smaller
perturbations. Again, the relations between this constitutive
property and the macroscopic behavior should be further explored in
future work. Bringing into consideration explicit macroscopic
measurements, one can constrain the various phenomenological
features of the theory of amorphous plasticity. This philosophy
provides a complementary approach to obtaining a better microscopic
understanding of the physical processes involved.

\section{Concluding Remarks}
\label{discussion}

We presented in this paper a detailed analysis of the linear stability of expanding
cavity modes in amorphous elasto-viscoplastic solids. The stability analysis is somewhat delicate due
to the non-stationarity of the problem, thus a perturbation may grow leaving the problem stable if this growth is slower than the growth of the radius of the cavity. The radial symmetry of the expanding cavity
makes it surprisingly resilient to perturbations in shape, velocity, external strains and pressure. On the other hand the radial symmetry may be lost due to perturbations in the internal state fields, especially $\chi$, and is also sensitive to details of the constitutive relations that are employed in the STZ theory. In this respect we highlight the role of the plastic rate of deformation dependent $\chi_\infty(\tau_0 \bar{D}^{pl})$ and of the stress-dependent rates of STZ transitions $\C C(s)$.
It is difficult to reach conclusive statements, since growth of perturbations beyond the linear order invalidate the approach taken here, calling for new algorithms involving surface tracking, where the size of perturbations is not limited. Nevertheless the results point out that instabilities are likely, motivating further research into the nonlinear regime. Of particular interest is the possibility to select particular forms of constitutive relations by comparing the predictions of the theory to macroscopic experiments. This appears as a promising approach in advancing the STZ theory towards a final form.

{\bf Acknowledgements} We thank T. Haxton and A. Liu for generously sharing with us their numerical data, and Chris Rycroft for pointing out an error in an early version of the manuscript. This work had been supported in part by the German Israeli Foundation and the Minerva Foundation, Munich, Germany. E. Bouchbinder acknowledges support from the Center for Complexity Science and the Lady Davis Trust.

\appendix

\section{Differential operators in polar coordinates}
\label{polar}

The aim of this Appendix is to derive some expressions in polar coordinates that were used earlier in the paper. Specifically, our goal is to calculate the divergence and covariant derivative of a tensor in polar coordinates. We represent a general second order tensor $\B T$ in polar coordinates as
\begin{equation}
\label{Tcomp}
\B T = T_{ij} \B e_i \otimes \B e_j \ ,
\end{equation}
where $\B e_i$ and $\B e_j$ are unit vectors in polar coordinates and $\otimes$ denotes a tensor product.
Using Eqs. (\ref{polarO})-(\ref{unit_vectors}) we obtain
\begin{eqnarray}
&&\B e_r \cdot \left(\B \nabla \cdot \B T \right)= \partial_r T_{rr}+\frac{T_{rr}}{r}+\frac{\partial_\theta T_{\theta r}}{r} -\frac{T_{\theta\theta}}{r} \ ,\nonumber\\
&&\B e_\theta \cdot \left(\B \nabla \cdot \B T\right)= \partial_r T_{r\theta}+\frac{T_{r\theta}}{r}+\frac{T_{\theta r}}{r}+ \frac{\partial_\theta T_{\theta \theta}}{r}\ .
\end{eqnarray}
Substituting Eqs. (\ref{sig_p_s}) for $\B T$, we obtain the right-hand-sides of Eqs. (\ref{EOM}).

We proceed now to calculate the covariant derivative of a tensor $\B v \cdot \B \nabla \B T$. Using Eqs. (\ref{polarO}), (\ref{unit_vectors}) and (\ref{Tcomp}) we obtain
\begin{eqnarray}
\left(\B v \cdot \B \nabla \B T\right)_{rr}&=&v_r \partial_r T_{rr}+\frac{v_\theta}{r}\partial_\theta T_{rr}-\frac{v_\theta}{r}T_{r\theta}-\frac{v_\theta}{r}T_{\theta r} \ ,\nonumber\\
\left(\B v \cdot \B \nabla \B T\right)_{r\theta}&=&v_r \partial_r T_{r\theta}+\frac{v_\theta}{r}T_{rr}+\frac{v_\theta}{r}\partial_\theta T_{r\theta}-\frac{v_\theta}{r}T_{\theta\theta} \ , \nonumber\\
\left(\B v \cdot \B \nabla \B T\right)_{\theta r}&=& v_r \partial_r T_{\theta r}+ \frac{v_\theta}{r} T_{rr}+\frac{v_\theta}{r}\partial_\theta T_{\theta r}-\frac{v_\theta}{r}T_{\theta\theta} \ ,\nonumber\\
\left(\B v \cdot \B \nabla \B T\right)_{\theta\theta}&=&v_r\partial_r T_{\theta\theta}+\frac{v_\theta}{r}T_{r \theta}+\frac{v_\theta}{r}T_{\theta r}+\frac{v_\theta}{r}\partial_\theta T_{\theta \theta} \ .\nonumber\\
\end{eqnarray}
Substituting Eqs. (\ref{eq:stress-strain}) for $\B T$ we obtain the needed expressions for $\left(\B v \cdot \B \nabla \B \epsilon^{el}\right)_{ij}$
in Eq. (\ref{eq:DA_ij}).

\section{The quasi-static and incompressible case}
\label{QS}

The aim of this appendix is to
derive independently the linear perturbation theory for a
quasi-static and incompressible case and to compare to the inertial and
compressible case in the limit of large bulk modulus $K$ and small velocities $v$. We show that the results in this limit agree, giving us some degree of confidence in the derivation and the numerical implementation of the
equations in both cases.

The unperturbed problem in the quasi-static and incompressible limit
was discussed in detail in \cite{07BLLP} and is obtained by
taking the quasi-static and the incompressible limits in the equations
of Sec. \ref{zeroth}. Before considering the linear stability problem, we stress that the linear perturbation theory of the STZ equations, presented in Sec. (\ref{perturbSTZ}) remains unchanged in the present analysis. Only the equations of motion and the kinematic equations are being modified. In the absence of inertial terms, the equations of motion (\ref{eqmot1})
become
\begin{equation}
\frac{\partial \tau}{\partial r} + \frac{1}{r} \frac{
\partial  s}{\partial \theta} - \frac{1}{r}
\frac{ \partial  p} {\partial \theta} +\frac{2 \tau}{r} = 0 \ ,
\end{equation}
\begin{equation}
\frac{1}{r}\frac{\partial \tau}{\partial \theta} -\frac{1}{r^2}
\frac{\partial }{\partial r} \left ( r^2 s \right) = \frac{\partial
p}{\partial r} \ .
\end{equation}
To first order we obtain
\begin{equation}\label{force1}
-\frac{1}{r}n \tau^{(1)}  - \frac {2 s^{(1)}}{r}  = \frac{\partial
p^{(1)}}{\partial r}  + \frac{\partial s^{(1)}}{\partial r} \ ,
\end{equation}
\begin{equation}\label{force2}
\frac{\partial \tau^{(1)}}{\partial r} + \frac{n }{r} (s^{(1)} -
p^{(1)}) +\frac{2 \tau^{(1)}}{r} = 0 \ .
\end{equation}

The boundary conditions of Eqs. (\ref{spb})-(\ref{taub}) can be further simplified by using the
force balance equation to zeroth order and the zeroth order boundary conditions of Eq. (\ref{BC0})
\begin{equation}
\label{use_eom} \frac{\partial p^{(0)}}{\partial r} =
-\frac{\partial s^{(0)}}{\partial r} - \frac{2s^{(0)}}{r} \ .
\end{equation}
Substituting into (\ref{spb}) and (\ref{taub}) we obtain
\begin{eqnarray}
\label{spb1} s^{(1)}(R^{(0)})\!+\!p^{(1)}(R^{(0)})\!\!&=&\!\!\frac{2s^{(0)}(R^{(0)})R^{(1)}}{R^{(0)}} \ ,\\
\label{taub1}\tau^{(1)}(R^{(0)}) \!\!&=&\!\! n
\frac{2s^{(0)}(R^{(0)})R^{(1)}}{R^{(0)}}.
\end{eqnarray}
In addition, all the first order fields decay as $r\!\to\!\infty$.
In principle, the initial conditions for the partial differential
equations for the first order fields depend on the type of
perturbation under consideration. For explicit perturbations in the
shape of the cavity, i.e. $R^{(1)}(0)\!\ne\!0$, we can determine the
initial stress field by assuming it is simply the quasi-static linear
elastic solution corresponding to the perturbed circle. In order to obtain this solution
we start with the bi-Laplace equation for the Airy stress function
$\Phi$ \cite{86LL}
\begin{equation}
\nabla^2\nabla^2\Phi=0 \label{bi-Laplace} \ ,
\end{equation}
where the stress tensor components are given by
\begin{eqnarray}
\sigma_{rr}&=&\frac{1}{r}\frac{\partial \Phi}{\partial r}+\frac{1}{r^2}\frac{\partial^2 \Phi}{\partial \theta^2} \ , \nonumber\\
\sigma_{\theta \theta}&=&\frac{\partial^2 \Phi}{\partial r^2} \
,\quad \sigma_{r \theta}=-\frac{\partial}{\partial
r}\left(\frac{1}{r}\frac{\partial \Phi}{\partial \theta} \right)
\label{Airy_components} \ .
\end{eqnarray}
We then expand the solution in the form
\begin{equation}
\Phi(r,\theta)=\Phi^{(0)}(r)+\Phi^{(1)}(r)e^{in\theta} \label{Phi} \
.
\end{equation}
The general solutions for $\Phi^{(1)}(r)$, that also decay at
infinity, are given by
\begin{equation}
\Phi^{(1)}(r) = a r^{-n+2}+ b r^{-n} \ ,
\end{equation}
with $n\!>\!0$.
Substituting in Eqs. (\ref{Airy_components}), using the boundary
conditions to first order and the following zeroth order solution
\begin{equation}
\sigma_{rr,\theta \theta}^{(0)} = \sigma^{\infty}\left(1 \mp
\frac{\left(R^{(0)}\right)^2}{r^2}\right)\ ,\quad
\sigma_{r\theta}^{(0)} = 0 \ ,
\end{equation}
one obtains
\begin{equation}
a=-\left[R^{(0)} \right]^n\!\!
\sigma^{\infty}\frac{R^{(1)}}{R^{(0)}}\ ,\quad b=\left[R^{(0)}
\right]^{n+2}\!\! \sigma^{\infty}\frac{R^{(1)}}{R^{(0)}}\ .
\end{equation}
The resulting stress components are easily calculated, from which we
obtain
\begin{eqnarray}
\!\!\!&p&\!\!\!^{(1)} =
2\sigma^\infty\frac{R^{(1)}}{R^{(0)}}(1-n)\left(\frac{R^{(0)}}{r}\right)^n\
,\quad s^{(1)} = \tau^{(1)} = \nonumber\\
\!\!\!&\sigma&\!\!\!^\infty\frac{R^{(1)}}{R^{(0)}}\left[n(1-n)\left(\frac{R^{(0)}}{r}\right)^n\!+\!n(n+1)\left(\frac{R^{(0)}}{r}\right)^{n+2}\right]\ .\nonumber\\
\end{eqnarray}
These are the initial conditions for the first order stress tensor components in terms of the initial
$R^{(1)}$. To proceed we expand Eqs. (\ref{el_pl}) to first order, assuming $K\!\to\!\infty$,
\begin{eqnarray}
\label{kinematic_1st_a} &&\frac{\partial v^{(1)}_r}{\partial
r}=\\
&& -\frac{1}{2 \mu} \left[\! \frac{\partial s^{(1)}}{\partial t}
\!+\! v_r^{(0)} \frac{\partial s^{(1)}}{\partial r} \!+\! v^{(1)}_r
\frac{\partial s^{(0)}}{\partial r}\right]
\!-\! {D^{pl}}^{(1)}\!,\nonumber\\
\label{kinematic_1st_b} &&\frac{\!-\!n v^{(1)}_\theta \!+\!
v^{(1)}_r}{r} = \\
&&\frac{1}{2 \mu} \left[\! \frac{\partial s^{(1)}}{\partial t} +
v_r^{(0)} \frac{\partial s^{(1)}}{\partial r} \!+\! v^{(1)}_r
\frac{\partial s^{(0)}}{\partial r}\right]
\!+\! {D^{pl}}^{(1)},\nonumber\\
\label{kinematic_1st_c} &&\frac{1}{2} \left[\!\frac {\partial
v^{(1)}_\theta}{\partial r} \!+ \!\frac{n v^{(1)}_r\!-\!
v^{(1)}_\theta}{r}
\right] = \\
&&\frac{1}{2 \mu} \left[\! \frac{\partial \tau^{(1)} }{\partial t} +
v^{(0)}_r \frac{\partial \tau^{(1)}}{\partial r}-\frac{2 s^{(0)} v_\theta^{(1)}}{r} \right] +
{D^{pl}_{r \theta}}^{(1)} \nonumber\ .
\end{eqnarray}
In order to propagate $s^{(1)}$ and $\tau^{(1)}$ in time according
to these equations we need to know $v^{(1)}_r$ and $v^{(1)}_\theta$
at each time step. However, a basic feature of the quasi-static
problem is that there is no evolution equation for the velocity
field. Therefore, we must calculate $v^{(1)}_r$ and $v^{(1)}_\theta$
in a different way.

We now discuss the major mathematical difficulty in the quasi-static
formulation, i.e. the absence of an explicit evolution equation for
the velocity field $\B v^{(1)}$. To overcome this difficulty, we
should derive new {\em ordinary differential equations} for
$v^{(1)}_r$ and $v^{(1)}_\theta$ such that their time evolution is
inherited from the other fields in the problem. The first equation can be
obtained readily by adding (\ref{kinematic_1st_a}) to
(\ref{kinematic_1st_b})
\begin{equation}
\label{final_1} \frac{\partial v^{(1)}_r}{\partial r}
+\frac{v_r^{(1)}}{r}-\frac{n v^{(1)}_\theta}{r}=0 \ ,
\end{equation}
from which we can extract $v^{(1)}_\theta$
\begin{equation}
\label{theta} v^{(1)}_\theta = \frac{1}{n}\left(r\frac{\partial
v^{(1)}_r}{\partial r} +v_r^{(1)}\right) \ .
\end{equation}
In order to obtain the second equation, we eliminate $p^{(1)}$ from
the equations by operating with $\frac{\partial }{\partial r}
\frac{r}{n}$ on Eq. (\ref{force2}), adding the result to Eq.
(\ref{force1}) and taking the partial time derivative to obtain
\begin{equation}
\label{elim_p} 2\frac{\partial \dot s^{(1)}}{\partial r} +
\frac{1}{n} \frac{\partial}{\partial r}\left(r\frac{\partial \dot
\tau^{(1)}}{\partial r}\right) +\frac{2}{n}\frac{\partial\dot
\tau^{(1)} }{\partial r} + \frac{n\dot \tau^{(1)}}{r}  + \frac{2
\dot s^{(1)}}{r}=0 \ .
\end{equation}
Here and elsewhere the dot denotes partial time derivative. Using
Eqs. (\ref{kinematic_1st_a}) and (\ref{kinematic_1st_c}) we obtain
\begin{eqnarray}
\label{tau_dot} \!\!\!\!&\dot\tau^{(1)}&\!\!\!=2\mu\!
\left[\!-\!{D^{pl}_{r \theta}}^{(1)}\!+\!\frac{1}{2}\left(\! \frac
{\partial v^{(1)}_\theta}{\partial r} + \frac{n v^{(1)}_r -
v^{(1)}_\theta}{r}
\right)\!\right]\nonumber\\
&&-v_r^{(0)}\frac{\partial\tau^{(1)}}{\partial
r}+\frac{2s^{(0)}v_\theta^{(1)}}{r}\ ,\\
\label{s_dot}\!\!\!\!&\dot s^{(1)}&=-\!2\mu\! \left(\!\frac{\partial
v_r^{(1)}}{\partial
r}\!+\!{D^{pl}}^{(1)}\right)-\!v_r^{(1)}\frac{\partial
s^{(0)}}{\partial r}\nonumber\\
&&-v_r^{(0)}\frac{\partial s^{(1)}}{\partial r} \ .
\end{eqnarray}
Substituting the last two relations in Eq. (\ref{elim_p}) and using Eq.
(\ref{theta}), we obtain a {\em fourth order linear ordinary
differential equation} for $v_r^{(1)}$. Since it is straightforward to obtain, but very lengthy, we do not write it explicitly here.
It is important to note that the coefficients in this equation
depend on time and therefore by solving it at each time step we
effectively have a time evolution for the velocity field. Once one
solves for $v_r^{(1)}$, Eq. (\ref{theta}) can be used to calculate
$v^{(1)}_{\theta}$. The forth order linear differential equation requires four boundary
conditions.

The first boundary condition is obtained by using Eq. (\ref{taub1}), with Eqs. (\ref{kinematic_1st_c}), (\ref{edge_velocity1}) and (\ref{theta}) we obtain a linear relation between $v_r^{(1)}(R^{(0)})$, $\partial_r v_r^{(1)}(R^{(0)})$ and $\partial^2_r
v_r^{(1)}(R^{(0)})$, which is the required boundary condition.
Another boundary condition is obtained by multiplying Eq.
(\ref{force2}) by r, operating with $\frac{{\cal D} }{{\cal D}
t}=\partial_t +v_r^{(0)} \partial_r$ on the result and using Eq. (\ref{spb1}). Additional simple manipulations result
in a linear relation between $v_r^{(1)}(R^{(0)})$, $\partial_r v_r^{(1)}(R^{(0)})$, $\partial^2_r
v_r^{(1)}(R^{(0)})$ and $\partial^3_r v_r^{(1)}(R^{(0)})$.
This is a second boundary relation. Two other boundary conditions
are obtained from the requirement that $v_r^{(1)}$ vanishes at
$\infty$ with vanishing derivative
\begin{equation}
\label{BCinfty} v_r^{(1)}(\infty)=0 \quad\hbox{and}\quad\
\frac{\partial v_r^{(1)}(\infty)}{\partial r}=0 \ .
\end{equation}
With these four boundary conditions the forth order differential equation can be solved in the following way: at each step we guess
$v_r^{(1)}(R^{(0)},t)$ and $\partial_r v_r^{(1)}(R^{(0)},t)$ and use
the first two boundary conditions to calculate $\partial_r^2
v_r^{(1)}(R^{(0)},t)$ and $\partial_r^3 v_r^{(1)}(R^{(0)},t)$. Then
we use the forth order differential equation to calculate $v_r^{(1)}$ and $\partial_r v_r^{(1)}$ at $\infty$. We improve our guess until the solution
satisfies Eqs. (\ref{BCinfty}) (the shooting method).

\begin{figure}
\centering \epsfig{width=.47\textwidth,file=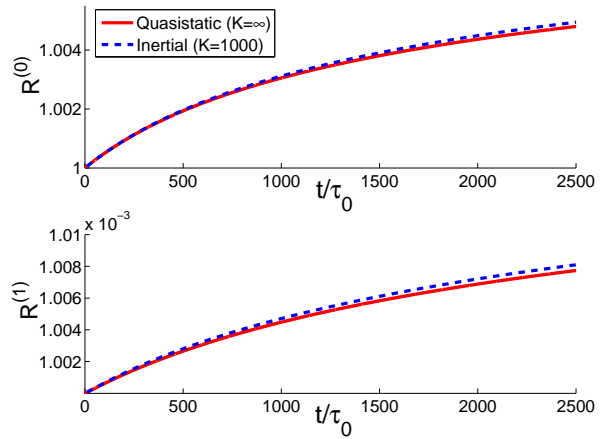}
\caption{(Color online) Upper panel: $R^{(0)}$ as a function of time
for both the quasi-static (solid line) and the inertial (dashed line)
formulations. We used $\sigma^\infty\!=\!2$, $\zeta\!=\!15$ and $K\!=\!1000$ in the
inertial case to ensure small velocities and almost
incompressibility. Lower panel: $R^{(1)}$ as a
function of time for both the quasi-static (solid line) and the
inertial (dashed line) formulations with a discrete wave-number
$n\!=\!2$. The agreement of the curves in both panels is favorable,
where the small differences are attributed to the finite bulk
modulus $K$ in the inertial case.}\label{compareR}
\end{figure}

\begin{figure}
\centering \epsfig{width=.47\textwidth,file=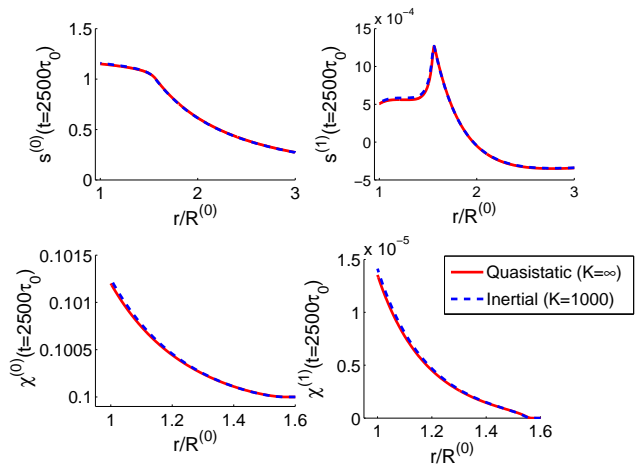}
\caption{(Color online) The fields $s^{(0)}(r)$, $s^{(1)}(r)$,
$\chi^{(0)}(r)$ and $\chi^{(1)}(r)$ at $t=2500\tau_0$ for both the
quasi-static (solid line) and the inertial (dashed line)
formulations. The parameters are the same as those of Fig. \ref{compareR}.}\label{compareFields}
\end{figure}

Thus, we have a complete solution procedure (assuming that the
plastic rate of deformation is known, see Section \ref{perturbSTZ});
for a given $s^{(1)}(r,t)$ and $\tau^{(1)}(r,t)$ we solve the forth order differential equation for
$v_r^{(1)}(r,t)$ following the procedure described above. Having
$v_r^{(1)}(r,t)$ we use Eq. (\ref{theta}) to obtain
$v_{\theta}^{(1)}(r,t)$. Then we use Eqs. (\ref{kinematic_1st_a}),
(\ref{kinematic_1st_c}) and (\ref{edge_velocity1}) to propagate
$s^{(1)}(r)$, $\tau^{(1)}(r)$ and $R^{(1)}$ in time. We follow the
same procedure at each time step to obtain the full time evolution
of the perturbation. We note that we have eliminated $p^{(1)}(r,t)$
from the problem, though we can calculate it at every time step
using Eq. (\ref{force1}) or (\ref{force2}).

We are now able to compare the quasi-static and incompressible case to the inertial and
compressible counterpart in the
limit of small velocities and large bulk modulus $K$.
We introduced at $t\!=\!0$ a perturbation of
magnitude $R^{(1)}/R^{(0)}\!=\!10^{-3}$ to the radius of the cavity,
with a discrete wave-number $n\!=\!2$, and solved the dynamics in
both formulations. We chose $\sigma^\infty\!<\!\sigma^{th}$ such
that the velocities are small and $K\!=\!1000$ in the inertial case
in order to approach the incompressible limit. In Fig.
\ref{compareR} we compare $R^{(0)}$ and $R^{(1)}$ for both the
quasi-static and the inertial formulations. The agreement is
good. Note that the stability of the expanding cavity depends
on the time dependence of the ratio $R^{(1)}/R^{(0)}$; however, we
do not discuss the stability yet, but focus on the comparison
between the two formulations. In Fig. \ref{compareFields} we further
compare the predictions of the two formulations for the zeroth and
first order deviatoric stress field $s$ and effective disorder
temperature $\chi$ at a given time. In all cases the differences are
practically indistinguishable. We thus conclude that the quasi-static
formulation and the inertial one agree with one another, giving us
some confidence in the validity of both. In particular we conclude
that the inertial formulation can be used with confidence also for
high velocities where the quasi-static counterpart becomes invalid.


\begin{thebibliography}{99}
\bibitem{comment0} Note that for situations where $\chi_\infty$ can become very large one should rewrite Eq. (\ref{eq:chi}) such that term in the square brackets reads $[1-\chi/\chi_\infty]$. This modification is not needed here as we do not consider such situations.
\bibitem{comment} Mainly in situations where all the tensors can be simultaneously diagonalized, resulting in effectively scalar equations.
\end{thebibliography}
\end{document}